\documentclass[sigconf]{acmart}

\def\BibTeX{{\rm B\kern-.05em{\sc i\kern-.025em b}\kern-.08emT\kern-.1667em\lower.7ex\hbox{E}\kern-.125emX}}

\newcommand{\bcl}{BCL}
\newcommand{\bclfullname}{Berkeley Container Library}
\newcommand{\container}{ObjectContainer}

\usepackage{url}
\usepackage{adjustbox}
\usepackage{listings}
\usepackage{booktabs}
\usepackage{amsmath,amsfonts}
\usepackage{algorithm}
\usepackage[noend]{algpseudocode}
\usepackage{graphicx}
\usepackage{textcomp}
\usepackage{color}
\usepackage{paralist}
\usepackage{enumitem}

\usepackage{tikz}

\makeatletter
\def\mdseries@tt{m}
\makeatother
\usepackage[frozencache]{minted}

\makeatletter
\def\BState{\State\hskip-\ALG@thistlm}
\makeatother

\setcopyright{none}
\renewcommand\footnotetextcopyrightpermission[1]{} 
\usetikzlibrary{shapes,arrows,shadows}
\usetikzlibrary{patterns}
\usetikzlibrary{positioning}

\usepackage{etoolbox}
\makeatletter
\patchcmd{\@makecaption}
  {\scshape}
  {}
  {}
  {}
\makeatother

\begin{document}

\title{\bcl{}: A Cross-Platform Distributed Data Structures Library\\
}

\author{Benjamin Brock, Ayd{\i}n Bulu\c{c}, Katherine Yelick}

\affiliation{
    \institution{University of California, Berkeley}
    \institution{Lawrence Berkeley National Laboratory}
}

\email{{brock,abuluc,yelick}@cs.berkeley.edu}


\begin{abstract}
One-sided communication is a useful paradigm for irregular parallel
applications, but most one-sided programming environments, including
MPI's one-sided interface and PGAS programming languages, lack
application-level libraries to support these applications.
We present the \bclfullname{}, a set of generic, cross-platform, high-performance data
structures for irregular applications, including queues, hash tables,
Bloom filters and more.  \bcl{} is
written in C++ using an internal DSL called the \bcl{} Core that provides
one-sided communication primitives such as remote get and remote put operations.
The \bcl{} Core has backends for MPI, OpenSHMEM, GASNet-EX, and UPC++, allowing
\bcl{} data structures to be used natively in programs written using any of these
programming environments.
Along with our internal DSL, we present the \bcl{} \container{} abstraction,
which allows \bcl{} data structures to transparently serialize complex data
types while maintaining efficiency for primitive types.
We also introduce the set of \bcl{} data structures and evaluate their
performance across a number of high-performance computing systems, demonstrating
that \bcl{} programs are competitive with hand-optimized code, even while hiding
many of the underlying details of message aggregation, serialization, and
synchronization.
\end{abstract}



\maketitle

\lstset{basicstyle=\ttfamily\small,breaklines=true}

\section{Introduction}

Writing parallel programs for supercomputers is notoriously difficult, particularly
when they have irregular control flow and complex data distribution; however,
high-level languages and libraries can make this easier.  A
number of languages have been developed for high-performance computing,
including several using the
\textit{Partitioned Global Address Space (PGAS)} model: Titanium, UPC,
Coarray Fortran, X10, and Chapel
\cite{yelick1998titanium,upc2005upc,numrich1998co,charles2005x10,weiland2007chapel,chamberlain2007parallel}.
These languages are especially well-suited to problems that require asynchronous
one-sided communication, or communication that takes place without a matching
receive operation or outside of a global collective.
However, PGAS languages lack the kind of high level libraries that exist in
other popular programming environments.  For example, high-performance
scientific simulations written in MPI can leverage a broad set of numerical
libraries for dense or sparse matrices, or for structured, unstructured, or
adaptive meshes.  PGAS languages can sometimes use those numerical libraries,
but lack the kind of data structures that are important in some of the
most irregular parallel~programs.

In this paper we describe a library, the \bclfullname{} (\bcl{}) that is
intended to support applications with irregular patterns of communication and
computation and data structures with asynchronous access, for example hash tables and
queues, that can be distributed across processes but manipulated independently by
each process.  \bcl{} is designed to provide a complementary set of abstractions
for data analytics problems, various types of search algorithms, and other
applications that do not easily fit a bulk-synchronous model.
\bcl{} is written in C++ and its data structures are designed
to be \textit{coordination free}, using one-sided communication primitives that
can be executed using RDMA hardware without requiring coordination with remote
CPUs.  In this way, \bcl{} is consistent with the spirit of PGAS languages, but
provides higher level operations such as \textit{insert} and \textit{find} in a
hash table,
rather than low-level remote read and write.   As in PGAS languages, \bcl{} data
structures live in a global address space and can be accessed by every process
in a parallel program.  \bcl{} data structures are also \textit{partitioned} to
ensure good locality whenever possible and allow for scalable implementations
across multiple nodes with physically disjoint memory.

\bcl{} is cross-platform, and is designed to be agnostic about the underlying
communication layer as
long as it provides one-sided communication primitives.   It runs on top of
MPI's one-sided communication primitives, OpenSHMEM, and GASNet-EX, all
of which provide direct access to low-level remote read and write
primitives to buffers in memory \cite{gerstenberger2014enabling, chapman2010introducing, bonachea2017gasnet}.
\bcl{} provides higher level abstractions than these communication layers, hiding many of the details of buffering,
aggregation, and synchronization from users that are specific to a given data structure.
\bcl{} also has an experimental UPC++ backend, allowing \bcl{} data structures
to be used inside another high-level programming environment.  \bcl{} uses a
high-level data serialization abstraction called \container{}s to allow the
storage of arbitrarily complex datatypes inside \bcl{} data structures.  \bcl{}
\container{}s use C++ compile-time type introspection to avoid introducing any
overhead in the common case that types are byte-copyable.



We present the design of \bcl{} with an initial set of data structures and operations.  We
then evaluate \bcl{}'s performance on
ISx, an integer sorting mini-application, Meraculous, a mini-application taken
from a
large-scale genomics application, and a collection of microbenchmarks examining
the performance of individual data structure operations.
We explain how \bcl{}'s data structures and
design decisions make developing high-performance implementations of these
benchmarks more straightforward and demonstrate that \bcl{} is able to match
or exceed the performance of both specialized, expert-tuned implementations as
well as general libraries across three different HPC systems.

\subsection{Contributions}
\begin{enumerate}[leftmargin=*]
  \item A distributed data structures library that is designed for
        high performance and portability by using a small set of core primitives
        that can be executed on four distributed memory backends
  \item The \bcl{} \container{} abstraction, which allows data structures to
        transparently handle serialization of complex types while maintaining
        high performance for simple types
  \item A distributed hash table implementation that supports fast insertion
        and lookup phases, dynamic message aggregation, and individual
        insert and find operations
  \item A distributed queue abstraction for many-to-many data exchanges performed
        without global synchronization
  \item A distributed Bloom filter which achieves fully atomic
        insertions using only one-sided operations
  \item A collection of distributed data structures that offer
        \textit{variable levels of atomicity} depending on the call context using
        an abstraction called concurrency promises
  \item A fast and portable implementation of the Meraculous benchmark built in \bcl{}
  \item An experimental analysis of irregular data structures across three different
        computing systems along with comparisons between \bcl{} and other standard
        implementations.
\end{enumerate}


\section{Background and High-Level Design}
Several approaches have been used to address programmability issues in
high-performance computing, including parallel languages like Chapel, template
metaprogramming libraries like UPC++, and embedded DSLs like STAPL.  These
environments provide core language abstractions that can boost productivity,
and some of them have sophisticated support for multidimensional arrays.  However,
none of these environments feature the kind of rich data structure libraries
that exist in sequential
programming enviroments like C++ or Java.  A particular need is for
distributed memory data structures that allow for nontrivial forms of
concurrent access that go beyond partitioned arrays in order to address the needs
of irregular applications.  These data structures tend to have more complicated
concurrency control and locality optimizations that go beyond tiling and ghost
regions.

Our goal is to build robust, reusable, high-level components to support these
irregular computational patterns while maintaining performance close to
hardware limits.  We aim to achieve this goal using the following design principles.

\paragraph{Low Cost for Abstraction} While \bcl{} offers data structures with
high-level primitives like hash table and queue insertions, these commands will be
compiled directly into a small number of one-sided remote memory operations.
Where hardware support is available, all primary data structure operations, such
as reads, writes, inserts, and finds, are executed purely in RDMA
\textit{without requiring coordination with remote CPUs}.
\paragraph{Portability} \bcl{} is cross-platform and can be used natively in
programs written in MPI, OpenSHMEM, GASNet-EX, and UPC++.
When programs only use \bcl{} data structures, users can pick
whichever backend's implementation is most optimized for their system and
network hardware.
\paragraph{Software Toolchain Complexity} \bcl{} is a \textit{header-only}
library, so users need only include the appropriate header files and compile
with a C++-14 compliant compiler to build a \bcl{} program.
\bcl{} data
structures can be used in part of an application without having to
re-write the whole application or include any new dependencies.

\section{\bcl{} Core}
The \bcl{} Core is the cross-platform internal DSL we use to implement \bcl{}
data structures.  It provides a high-level PGAS memory model based on global
pointers, which are C++ objects that allow the manipulation of remote memory.
During initialization,
each process creates a \textit{shared memory segment} of a fixed size, and every
process can read and write to any location within another node's shared segment
using global pointers.  A global pointer is a C++ object that contains (1) the rank
of the process on which the memory is located and (2) the particular offset within
that process' shared memory segment that is being referenced.
Together, these two values uniquely identify a global
memory address.  Global pointers are regular data objects and can be passed
around between \bcl{} processes using communication primitives
or stored in global memory.  Global pointers
support pointer arithmetic operations analogous to local pointer arithmetic.

Global pointers support \textit{remote get} and \textit{remote put} operations,
which can be performed using an explicit \texttt{BCL::memcpy} operation or
with

\subsection{Communication Primitives}
\subsubsection{Writing and Reading}
The \bcl{} Core's primary memory operations involve writing and reading to
global pointers.  \textit{Remote get} operations read from a global pointer and
copy the result into local memory, and \textit{remote put} operations write
the contents of an object in local memory to a shared memory location referenced
by a global pointer.  Remote completion of put operations is not guaranteed
until after a memory fence such as a flush or barrier.

\begin{figure}
  \definecolor{BerkeleyGold}{HTML}{C4820E}
\definecolor{BerkeleyBlue}{HTML}{6c94cc}
\definecolor{BerkeleyGray}{HTML}{EEEEEE}

\pgfdeclarelayer{background}
\pgfdeclarelayer{foreground}
\pgfsetlayers{background,main,foreground}


\begin{tikzpicture}

\tikzstyle{backend}=[draw, fill=BerkeleyBlue, text width=8em,
    text centered, minimum height=2.5em,drop shadow]

\tikzstyle{dsl} = [backend, text width=10em, fill=BerkeleyGray,
    minimum height=6em, rounded corners, drop shadow]
\tikzstyle{library}=[draw, fill=BerkeleyGold, text width=8em,
    text centered, rounded corners, minimum height=2.5em,drop shadow]

\def\blockdist{2.3}
\def\edgedist{2.5}
  \begin{scope}[scale=0.7, transform shape]
    \node (wa) [dsl]  {Internal DSL (BCL Core)};
    \path (wa.west)+(-3.2,1.5) node (asr1) [backend] {MPI};
    \path (wa.west)+(-3.2,0.5) node (asr2)[backend] {OpenSHMEM};
    \path (wa.west)+(-3.2,-0.5) node (asr3)[backend] {GASNet-EX};
    \path (wa.west)+(-3.2,-1.5) node (asr4)[backend] {UPC++};

    \path (wa.east)+(\blockdist,0) node (vote) [library] {BCL Containers};

    \path [draw, ->] (asr1.east) -- node [above] {}
        (wa.160) ;
    \path [draw, ->] (asr2.east) -- node [above] {}
        (wa.172);
    \path [draw, ->] (asr3.east) -- node [above] {}
        (wa.184);
    \path [draw, ->] (asr4.east) -- node [above] {}
        (wa.200);
    \path [draw, ->] (wa.east) -- node [above] {}
        (vote.west);


  \end{scope}

\end{tikzpicture}
  \centering
  \caption{Organizational diagram of \bcl{}.}
  \label{fig:org-diagram}
  \vspace{-1em}
\end{figure}
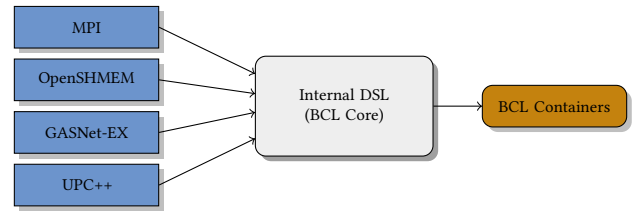

\subsubsection{Collectives}
\bcl{} includes the broadcast and allreduce collectives.
Depending on the backend, these may be implemented using raw remote put and
remote get operations, or, more likely, may map directly to high-performance
implementations offered by the backend communication framework.  In the work
presented here, collective performance is not critical, as they are mainly used
for transporting pointers and control values.

\subsubsection{Atomics}
\bcl{}'s data structures avoid coordination between CPUs, instead relying on
\textit{remote memory atomics} to maintain consistency.  \bcl{} backends must
implement at least the atomic compare-and-swap operation, since all other
atomic memory operations (AMOs) can be implemented on top of compare-and-swap \cite{herlihy1991wait}.
However, backends will achieve much higher performance by directly including any
atomic operations available in hardware.  Other atomic operations provided by
current \bcl{} backends and utilized by \bcl{} data structures include atomic
fetch-and-add and atomic-fetch-and-or.  We depend on backends to provide high
quality interfaces
to atomic operations as implemented in hardware, but also to provide atomic
operation support through active messages or progress threads when hardware
atomics are not available.

\subsubsection{Barriers}
\bcl{} applications enforce synchronization using \bcl{} barriers, which are both
barriers and memory fences, forcing ordering of remote memory operations.
In order for a rank to enter a barrier, all of its
memory operations must complete, both locally and at the remote target.  In order
for a rank to exit a barrier, all other threads must have entered the barrier.

\section{Parallel Patterns in \bcl}


When choosing data structures to implement in \bcl, we wanted to focus
on data structures that could exploit particular high-level parallel patterns~\cite{mattson2004patterns, mccool2012structured}.
While \bcl{} also efficiently supports commonly known data structure patterns
such as the Distributed Array Pattern~\cite{mattson2004patterns}, its novelty lies in its
support for more challenging irregular data access patterns as first-class citizens.
In particular, we chose to focus on exposing high-level data structures that
exploit two parallel patterns: (1) fine-grained, low-latency reads and writes,
and (2) asynchronous many-to-many redistribution of data. These patterns occur
in many applications that perform concurrent reads and writes in an unpredictable manner,
with prime examples coming from graph algorithms, computational chemistry, and bioinformatics.
These patterns can also be used in loosely synchronous applications~\cite{fox2014parallel} that require data
redistribution due to changes in the computational
structure as the algorithms proceed~\cite{ou1996fast}.

\subsection{Fine-Grained RDMA Operations}
For the first pattern,
we wanted to provide high-level interfaces for fine-grained operations that are
potentially complex, such as hash table operations, but in many cases will be
executed as a single RDMA operation.  For these low-latency operations, designing
a low-cost, header-only library where user code is compiled down to a small
number of calls to a backend library is essential to achieve performance.
Also essential to achieving performance for low-latency operations across a
variety of computing platforms is supporting multiple backends, since oftentimes
the best communication backend varies across supercomputing platforms.  Examples
of data structures we implemented which expose this pattern include hash tables
and Bloom filters, discussed in Sections~\ref{sec:hashtable}~and~\ref{sec:bloomfilter}.

\subsection{Many-to-Many Data Redistribution}
For the second pattern, we are interested in applications where each process wishes
to push data to other processes in an asynchronous, arbitrary manner.  MPI
all-to-all provides a restricted implementation of this pattern, where each process
gathers its data to be sent to each other process, then all processes take part
in a bulk synchronous all-to-all operation.  While there are asynchronous versions
of MPI all-to-all, it still restricts processes from generating new data after
the all-to-all operation has started, thus limiting the possibility for overlap
between communication and computation.  Sometimes this pattern is explicitly
present, such as in sorting or histogramming, but sometimes it can be exposed
by buffering and aggregating fine-grained operations.  In this paper, we first
build queue data structures (Section \ref{sec:queues}) that allow for arbitrary
data redistribution using
asynchronous queue insertions. Then, we design a ``hash table buffer''
data structure (Section \ref{sec:hashmapbuffer}) that allows users to buffer and
aggregate hash table insertions
transparently, transforming fine-grained, latency-bound operations into bulk,
bandwidth-bound ones.

\section{\bcl{} Data Structures}
\bcl{} data structures are split into two categories: \textit{distributed} and
\textit{hosted}.  Distributed data structures live in globally addressable memory
and are automatically distributed among all the ranks in a \bcl{} program.
Hosted data structures, while resident in globally addressable memory, are
hosted only on a particular process.  All other processes may read or write from
the data structure lying on the host process.  We have found hosted data structures
to be an important building block in creating distributed data structures.

All \bcl{} data structures are \textit{coordination free}, by which we mean
that primary data structure operations, such as insertions, deletions, updates,
reads, and writes, can be performed without coordinating with the CPUs of other
nodes, but purely in RDMA where hardware support is available.  Other operations,
such as resizing or migrating hosted data structures from one node to another,
may require coordination.  In particular, operations which modify the
size and location of the data portions of \bcl{} data structures must be
performed collectively, on both distributed and hosted data structures.  This
is because coordination-free data structure methods, such as insertions, use
global knowledge of the size and location of the data portion of the data
structure.  For example, one process cannot change the size or location of a
hash table without alerting other processes, since they may try to insert into
the old hash table memory locations.
Tables \ref{table:containers} and \ref{table:methods} give an overview of the
available data structures and operations.
Table \ref{table:methods} also gives the best-case cost of each operation in terms of remote reads $R$,
remote writes $W$, atomic operations $A$, local operations $\ell$, and global
barriers $B$.  As demonstrated by the table, each high-level data structure
operation is compiled down to a small number of remote memory operations.

All \bcl{} data structures are also generic, meaning they can be used to hold
any type, including complex, user-defined types.  Most common types will be
handled automatically, without any intervention by the user.  See Section~\ref{sec:container}
for a detailed description of \bcl{}'s lightweight serialization mechanism.

Many distributed data structure operations have multiple possible implementations
that offer varying levels of atomicity.  Depending on the context of a particular
callsite, only some of these implementations may be valid.  We formalize a
mechanism, called \textit{concurrency promises}, that allows users to optionally assert
invariants about a callsite context.  This allows \bcl{} data structures to use
optimized implementations that offer fewer atomicity guarantees when a user
guarantees that this is possible.  This mechanism is discussed in Section~\ref{sec:conprom}.

\begin{table}
  \centering
  \resizebox{\columnwidth}{!}{
  \begin{tabular}{| l | l | l |}
    \hline
      Data Structure & Locality & Description\\
      \hline
      \texttt{\bcl{}::HashMap} & Distributed & Hash Table\\
      \texttt{\bcl{}::CircularQueue} & Hosted & Multiple Reader/Writer Queue\\
      \texttt{\bcl{}::FastQueue} & Hosted & Multi-Reader \textit{or} Multi-Writer Queue\\
      \texttt{\bcl{}::HashMapBuffer} & Distributed & Aggregate hash table insertions\\
      \texttt{\bcl{}::BloomFilter} & Distributed & Distributed Bloom filter\\
      \texttt{\bcl{}::DArray} & Distributed & 1-D Array\\
      \texttt{\bcl{}::Array} & Hosted & 1-D Array on one process\\
    \hline
  \end{tabular}
  }
    \vspace{1em}
  \caption{\textsc{A summary of \bcl{} data structures.}}
  \label{table:containers}
  \vspace{-1em}
\end{table}

\begin{table*}
  \begin{adjustbox}{width=1.05\textwidth,center}
    \begin{tabular}{| l | l | l | l | l | l |}
      \hline
        Data Structure & Method & Collective & Description & Cost\\
        \hline
        \textbf{\texttt{\bcl{}::HashMap}}\\
        \hline
        & \texttt{bool insert(const K \&key, const V \&val)} & N &  Insert item into hash table. & 2A + W\\
        & \texttt{bool find(const K \&key, V \&val)} & N & Find item in table, return val. & 2A + R\\
        \hline
        \textbf{\texttt{\bcl{}::BloomFilter}}\\
        \hline
        & \texttt{bool insert(const T \&val)} & N & Insert item into Bloom filter, return true if already present. & $A$\\
        & \texttt{bool find(const T \&val)} & N & Find item in table, return whether present. & $R$\\
        \hline
        \textbf{\texttt{\bcl{}::CircularQueue}}\\
        \hline
        & \texttt{bool push(const T \&val)} & N & Insert item into queue. & $2A + W$\\
        & \texttt{bool pop(T \&val)} & N & Pop item into queue. & $2A + R$\\
        & \texttt{bool push(const std::vector <T> \&vals)} & N & Insert items into queue. & $2A + nW$\\
        & \texttt{bool pop(std::vector <T> \&vals, size\_t n)} & N & Pop items from queue. & $2A + nR$\\
        & \texttt{bool local\_nonatomic\_pop(T \&val)} & N & Nonatomically pop item from a local queue. & $\ell$\\
        & \texttt{void resize(size\_t n)} & Y & Resize queue. & $B + \ell$\\
        & \texttt{void migrate(size\_t n)} & Y & Migrate queue to new host. & $B + nW$\\
        \hline
        \textbf{\texttt{\bcl{}::FastQueue}}\\
        \hline
        & \texttt{bool push(const T \&val)} & N & Insert item into queue. & $A + W$\\
        & \texttt{bool pop(T \&val)} & N & Pop item into queue. & $A + R$\\
        & \texttt{bool push(const std::vector <T> \&vals)} & N & Insert items into queue. & $A + nW$\\
        & \texttt{bool pop(std::vector <T> \&vals, size\_t n)} & N & Pop items from queue. & $A + nR$\\
        & \texttt{bool local\_nonatomic\_pop(T \&val)} & N & Nonatomically pop item from a local queue. & $\ell$\\
        & \texttt{void resize(size\_t n)} & Y & Resize queue. & $B + \ell$\\
        & \texttt{void migrate(size\_t n)} & Y & Migrate queue to new host. & $B + nW$\\
        \hline
      \end{tabular}
  \end{adjustbox}
  \vspace{1em}
  \caption{A selection of methods from \bcl{} data structures.
  Costs are best case, using implementation chosen with no concurrency promises.
  $R$ is the cost of a remote read, $W$ the cost of a remote write, $A$ the cost
  of a remote atomic memory operation,
  $B$ the cost of a barrier, $\ell$ the cost of a local memory operation, and $n$
  the number of elements involved.}
  \label{table:methods}
  \vspace{-1em}
\end{table*}

\begin{table}
    \begin{tabular}{| l | l | l | l | l | l |}
      \hline
        Method & Concurrency Promise & Description & Cost\\
        \hline
        \textbf{\texttt{insert}}\\
        \hline
        \multicolumn{1}{|r|}{\footnotesize (a)} & \texttt{find | insert}& Fully Atomic & $2A + W$\\
        \multicolumn{1}{|r|}{\footnotesize (b)} & \texttt{local} & Local Insert & $\ell$\\
        \hline
        \textbf{\texttt{find}}\\
        \hline
        \multicolumn{1}{|r|}{\footnotesize (c)} & \texttt{find | insert}& Fully Atomic & $2A + R$\\
        \multicolumn{1}{|r|}{\footnotesize (d)} & \texttt{find} & Only Finds & $R$\\
        \hline
      \end{tabular}

  \vspace{1em}
  \caption{Implementations of data structure operations for \bcl{}'s hash table
           data structure.}
  \label{table:conpromhash}
  \vspace{-2em}
\end{table}

\begin{table}
    \begin{tabular}{| l | l | l | l | l | l |}
      \hline
        Method & Concurrency Promise & Description & Cost\\
        \hline
        \textbf{\texttt{push}}\\
        \hline
        \multicolumn{1}{|r|}{\footnotesize (a)} & \texttt{push | pop}& Fully Atomic & $2A + W$\\
        \multicolumn{1}{|r|}{\footnotesize (b)} & \texttt{push} & Only Pushes & $2A + W$\\
        \multicolumn{1}{|r|}{\footnotesize (c)} & \texttt{local} & Local Push & $\ell$\\
        \hline
        \textbf{\texttt{pop}}\\
        \hline
        \multicolumn{1}{|r|}{\footnotesize (d)} & \texttt{push | pop}& Fully Atomic & $2A + R$\\
        \multicolumn{1}{|r|}{\footnotesize (e)} & \texttt{pop} & Only Pops & $2A + R$\\
        \multicolumn{1}{|r|}{\footnotesize (f)} & \texttt{local} & Local Pop & $\ell$\\
        \hline
      \end{tabular}

  \vspace{1em}
  \caption{Implementations of data structure operations for \bcl{}'s circular
           queue data structure.}
  \label{table:conpromqueue}
  \vspace{-2em}
\end{table}

\subsection{Hash Table}
\label{sec:hashtable}
\bcl{}'s hash table is implemented as a
single logically contiguous array of hash table buckets distributed block-wise
among all processes.  Each bucket is a struct including a key, value, and status
flag.  Our hash table uses open addressing with quadratic probing to resolve hash
collisions.
As a result, neither insert nor find operations to our
hash table require any coordination with remote ranks.  Where hardware support
is available, hash table operations will take place purely with RDMA operations.
\subsubsection{Interface}
\bcl{}'s \texttt{\bcl{}::HashMap} is a distributed data structure.  Users can create
a \texttt{\bcl{}::HashMap} by calling the constructor as a collective operation
among all ranks.  \bcl{} hash tables are created with a fixed key and value type
as well as a fixed size.  \bcl{} hash tables use \container{}s, discussed in
Section~\ref{sec:container}, to store keys and values
of any type.  \bcl{} hash tables also use the standard C++ STL method for handling hash
functions, which is to look for a \texttt{std::hash~<K>} template struct in the
standard namespace that provides a mechanism for hashing key objects.

The hash table supports two primary methods, \texttt{insert} and
\texttt{find}.  Section \ref{sec:eval} gives a
performance analysis of our hash table.

\subsubsection{Atomicity}
By default, hash table insert and find operations are atomic with respect to one
another, including simultaneous insert operations and find operations using the
same key.
In addition to this default level of atomicity, users can pass a
concurrency promise as an optional argument at each callsite that can allow the
data structure to select a more optimized implementation with less strict
atomicity guarantees.  All the available implementations for insert and
find operations are shown in Table \ref{table:conpromhash}.

Our hash table uses a lightweight, per-bucket locking scheme.  Each hash table
bucket has a 32-bit \textit{used} flag that ensures atomicity of
operations.  The lowest 2 bits of this flag indicate the reservation status of
the bucket.  There are three possible states: (1) free, (2) reserved,
and (3) ready.  The free state represents an unused bucket, the reserved state
represents a bucket that has been reserved and will immediately be written to
by a process, and the ready state indicates that a bucket is ready to be read.
The
remaining 30 bits are \textit{read flag} bits, and they indicate, if flipped,
that a process is currently reading the hash table entry, and prevent another
process from writing to the entry before the other process has finished~reading.

\subsubsection{Insert Operations}
The default, fully atomic process for inserting (Table \ref{table:conpromhash}a)
requires two atomic memory operations
(AMOs) and a remote put with a flush.  First, the inserting process computes the
appropriate bucket.  Then it uses a compare-and-swap (CAS) operation to
set the bucket's status to reserved, a remote put to write
the correct key and value to the reserved bucket, followed by a flush
to ensure completion of the put, then finally an atomic XOR to set
the status of the bucket to ready.  Pseudocode for this implementation follows.

\begin{enumerate}
\item CAS bucket state from free $\rightarrow$ reserved.  If state is
      discovered to be reserved/ready, instead CAS ready $\rightarrow$ reserved.
\item Once successful, if the previous state was free, write the key and value
      to the bucket.  If the previous state was ready, check the key.  If key
      is incorrect, reset status, move to the next bucket, and begin at (1).
\item Set status bits to ready using an atomic XOR.  (Flipped read bits from
      attempted reads could interfere with a CAS, which is unnecessary.)
\end{enumerate}

In some special cases, we may wish to have processes perform local insertions
into their own portions of the hash table.  This may be done with only local
CPU instructions, not involving the NIC.  Crucially, this cannot be done when
other operations, such as general find or insert operations, might be executed,
since CPU atomics are not atomic with respect to NIC atomics.  This implementation
requires the concurrency promise \texttt{local} (Table \ref{table:conpromhash}b).

\subsubsection{Find Operations}
The default, fully atomic implementation of the find operation (Table \ref{table:conpromhash}c)
again involves two
AMOs and a remote read.  First, the process uses a fetch-and-or to set a random
read bit.  This keeps other processes from writing to the hash bucket before the process
has finished reading it.  Then, it reads the value, and, after reading, unsets
the read bit.  Pseudocode for this implementation follows.

\begin{enumerate}
\item Atomic fetch-and-or to set a random read bit. If bit was previously set,
      start again at (1).  If bit was unset and status was reserved, atomically
      fetch until status is ready.
\item Once read bit is set and status is ready, read the key and value.
\item Unset read bit using atomic fetch-and-and.  If key was incorrect, move on
      to next bucket.
\end{enumerate}

In the common case of a \textit{traversal phase} of an application, where no
insert operations may occur concurrent with find operations, we may use an
alternate implementation that requires no atomic operations (Table \ref{table:conpromhash}d),
but just a single
read operation to read the whole bucket including the reserved flag, key, and value.

\subsubsection{Hash Table Size}
A current limitation of \bcl{} is that, since hash tables are initialized to a
fixed size and do not dynamically resize, an insertion may fail.
In the future, we plan to support a dynamically resizing
hash table.  Currently, the user must call the collective \texttt{resize}
method herself when the hash table becomes full.

\subsection{Queues}
\label{sec:queues}
\bcl{} includes two types of queues: one, \texttt{CircularQueue}, is a general
multi-reader, multi-writer queue which supports variable levels of atomicity.
The second, \texttt{FastQueue}, supports multiple readers \textit{or} multiple
writers, but requires that read and write phases be
separated by a barrier.  Both queues are implemented as ring buffers and
are initialized with a fixed
size as a \textit{hosted} data structure, so while a queue is globally visible,
it is resident on only one process at a time.

\subsubsection{\texttt{FastQueue}}
\begin{figure}

\centering
\begin{tikzpicture}[>=latex,every node/.style={minimum width=1cm,minimum height=1.5em,outer sep=0pt,draw=black,fill=white,semithick,font=\footnotesize}]
  \tikzset{Filled/.style={minimum width=1cm,minimum height=1.5em,outer sep=0pt,draw=black,fill=white,semithick,pattern=north west lines}}
  \tikzset{Reserved/.style={minimum width=1cm,minimum height=1.5em,outer sep=0pt,draw=black,fill=white,semithick,pattern=dots}}
  \tikzset{Order/.style={circle, minimum size=1pt, outer sep=0pt,draw=black,fill=white,semithick,font=\footnotesize}}
  \tikzset{fontscale/.style = {draw=none,fill=none,font=\footnotesize}}
        \node at (0,0) (A) {};
        \node [anchor=north] at (A.south) (B) {};
        \node [Filled, anchor=north] at (B.south) (C) {};
        \node [Filled, anchor=north] at (C.south) (D) {};
        \node [Filled, anchor=north] at (D.south) (E) {};
        \node [Reserved, anchor=north] at (E.south) (F) {};
        \node [Reserved, anchor=north] at (F.south) (G) {};

        \node [fontscale,left = 2em of C] (Head) {Head};
        \draw [->] (Head) -- (C);

        \node [fontscale,left = 2em of E] (Tail) {Tail};
        \draw [->] (Tail) -- (E);

        \node [fontscale,left = 2em of G] (Ntail) {New Tail};
        \draw [->] (Ntail) -- (G);

        \draw [->] (Tail) -- (Tail |- Ntail.north) node [midway, left, draw=none, fill=none] (FAD) {\textbf{\texttt{fetch\_and\_add}}};

        \node [fontscale,right = 2em of F] (rput) {\textbf{\texttt{rput}}};
        \draw [->] (rput) -> (F);
        \draw [->] (rput) -> (G);

        \node [Order, above = 0.26em of FAD, scale=0.8] (1) {1};
        \node [Order, above = 0.26em of rput, scale=0.8] (2) {2};

        \matrix [draw,below right] at (current bounding box.north east) {
          \node [Filled, label=right:Filled] {}; \\
          \node [Reserved, label=right:Reserved] {}; \\
        };

\end{tikzpicture}
\caption{Process for pushing values to a \bcl{} \texttt{FastQueue}.  First (1) a \texttt{fetch\_and\_add}
         operation is performed, which returns a reserved location where values
         can be inserted.  Then (2) the values to be inserted are copied to the
         queue.}
\label{fig:queue_insert}
\vspace{-1em}
\end{figure}
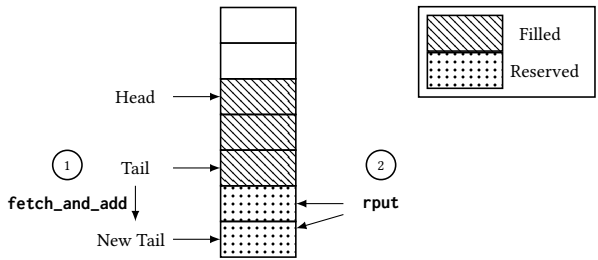

\texttt{FastQueue} uses three shared objects: a data segment, where
queue elements are stored; a shared integer that stores the head of the queue; and a
shared integer that stores the tail of the queue.
To insert, a process first increments the tail using an atomic
fetch-and-add operation, checks that this does not surpass the
head pointer, and then inserts its value or values into the data segment of the
queue.  An illustration of a push operation is shown in Figure \ref{fig:queue_insert}.
In general, the head overrun check is performed without a remote
memory operation by caching the position of the head pointer, so an insertion
requires only two remote memory operations.  We similarly cache the location of the
tail pointer, so pops to the queue usually require only one atomic memory operation
to increment the head pointer and one remote memory operation to read the popped
values.

\subsubsection{\texttt{CircularQueue}}
To support concurrent reads and writes, \texttt{CircularQueue} has an additional
set of head and tail pointers which indicate which portions of data in the
queue are ready to be read.  There are multiple implementations of push and pop
for a \texttt{CircularQueue} data structure, as listed in Table~\ref{table:conpromqueue}.

\subsubsection{Push and Pop Operations}

The default fully atomic implementation used for insertion (Table~\ref{table:conpromqueue}a)
into a
\texttt{CircularQueue} data structure involves 2 atomic operations and a remote
put operation with a flush.  First, we issue a fetch-and-add operation to
increment the tail pointer, then write the data to the queue and flush it.
Finally, we must perform a compare-and-swap operation to increment the ``tail ready''
pointer, indicating that the pushed data is ready to be read.  A CAS is necessary
for the
final step because a fetch-and-add could increment the ready pointer to mistakenly
mark other processes' writes as ready to be read.  In the case where no pop
operations will be performed before a barrier, we may perform the final atomic
increment using a fetch-and-add (Table~\ref{table:conpromqueue}b).
An analogous implementation is used for pop operations
(Table~\ref{table:conpromqueue}d~and~\ref{table:conpromqueue}e).

Both queues support resizing as well as migrating to another host process, both
as collective operations.
We evaluate the performance of our circular queue data structures in Section \ref{sec:cqeval}.

\subsubsection{Advantage of FastQueue}
\texttt{FastQueue} has the advantage of requiring one fewer AMO per push or pop.
While the \texttt{CircularQueue} does support variable levels of atomicity, allowing
the final pop to be a single non-blocking fetch-and-add operation, we felt that
this was an important enough overhead to warrant a separate version of the data
structure, since queues that support only multi-reader and multi-writer \textit{phases}
are crucial to several of the algorithms that we explored.

%

\subsection{Buffering Hash Table Insertions}
\label{sec:hashmapbuffer}
Many applications, such as the Meraculous benchmark, exhibit \textit{phasal}
behavior, where there is an insert phase, followed by a barrier, followed by
a read phase.  We anticipate that this is likely to be a common case, and
so have created a hash table \textit{buffer} data structure that accelerates hash table
insertion phases.  An application programmer can create a new \texttt{\bcl{}::HashMapBuffer}
on top of an existing hash table. The user then inserts
directly into the hash map buffer object using the same methods provided by the
hash table.  This simple code transformation is demonstrated in Figure \ref{fig:hashbuff}.
While the hash table interface ensures ordering of hash table insertions,
insertions into the hash table buffer are non-blocking, and ordering is no longer
guaranteed until after an explicit flush operation.
The hash table buffer implementation creates a \texttt{FastQueue} on each
node as well as local buffers for each other node.  When a user inserts into the
hash table buffer, the insert will be stored in a buffer until the buffer reaches
its maximum size, when it will be pushed to the queue lying on the appropriate
node to be staged for insertion.
At the end of an insert phase, the user calls the \texttt{flush()}
method to force all buffered insertions to complete.  Insertions into the actual
table will be completed using a local, fast hash table insertion (Table~\ref{table:conpromhash}b).
The hash map buffer
results in a significant performance boost for phasal applications, as discussed
in Section~\ref{sec:hasheval}.

\subsection{Bloom Filters}
\label{sec:bloomfilter}
A Bloom filter is a space-efficient, probabilistic data structure that answers
queries about set membership \cite{bloom1970space}.  Bloom filters can be used
to improve the
efficiency of hash tables, sets, and other key-based data structures.
Bloom filters support two operations, \textit{insert} and \textit{find}.
To insert a value into the Bloom filter, we use $k$ hash functions
to hash the value to be inserted $k$ times, resulting in $k$ locations in a bit
array that will all be set to one.  To check if a value is present in a Bloom
filter, the value is hashed $k$ times, and if each of the corresponding bits is
set, the value is said to be present.  Because of hash collisions, a Bloom
filter may return false positives, although it will never return false
negatives.

\subsubsection{Distributed Bloom Filter}
A simple Bloom filter can be implemented in distributed memory as an array of
integers distributed block-wise across all processes.
To insert a value into this Bloom filter, the value to be inserted is hashed $k$
times to determine which bits in the distributed Bloom filter must be set.
Then, the appropriate bits are set using
atomic fetch-and-or operations to the corresponding integers.
To check whether a value is in the Bloom filter, the value can be hashed $k$ times
and the corresponding locations in the array checked.

There is, however, a fundamental limitation in implementing a distributed Bloom filter
this way, as it results in a loss of atomicity for insert operations.
For many applications, it may be useful to have an atomic
insertion operation that inserts a value into a Bloom filter and atomically returns a
boolean value indicating whether the value was already present in the Bloom
filter.  In a regular serial Bloom filter, we define a value as already present
if all $k$ bits for our value were already set in the filter, and not already
present if we were required to flip any bits in the filter.  Since our
insertion operation consists of multiple AMOs which cannot be guaranteed to be
executed together atomically, we cannot guarantee that two processes which
attempt to insert the same value simultaneously into the Bloom filter will not
both believe that they were the first process to insert that value into the
Bloom filter.  This violates the invariant that a Bloom filter will return no
false negatives, so the Bloom filter described above cannot provide this
information.

There is also a disadvantage in terms of communication cost for this
implementation, since performing an insert requires flipping $k$ bits,
generally resulting in $k$
independent atomic operations.

\begin{figure}
\begin{minted}[mathescape,
               linenos,
               numbersep=5pt,
               gobble=2,
               frame=lines,
               framesep=2mm,
               fontsize=\small,
               escapeinside=||]{C++}
auto sort(const std::vector<int>& data) {
  std::vector<std::vector<int>> buffers(|\bcl|::nprocs());

  std::vector<|\bcl|::FastQueue<int>> queues;
  for (size_t rank = 0; rank < |\bcl|::nprocs(); rank++) {
    queues.push_back(|\bcl|::FastQueue<int>(rank, queue_size));
  }

  for (auto& val : data) {
    size_t rank = map_to_rank(val);
    buffers[rank].push_back(val);
    if (buffers[rank].size() >= message_size) {
      queues[rank].push(buffers[rank]);
      buffers[rank].clear();
    }
  }

  for (size_t i = 0; i < buffers.size(); i++) {
    queues[i].push(buffers[i]);
  }

  |\bcl|::barrier();

  std::sort(queues[|\bcl|::rank()].begin().local(),
            queues[|\bcl|::rank()].end().local());
  return queues[|\bcl|::rank()].as_vector();
}
\end{minted}
\vspace{-0.5em}
  \caption{Our bucket sort implementation in \bcl{} for the ISx benchmark.}
  \label{fig:sortex}
  \vspace{-1em}
\end{figure}

\begin{figure}
\begin{minted}[mathescape,
               linenos,
               numbersep=5pt,
               gobble=2,
               frame=lines,
               framesep=2mm,
               fontsize=\small,
               escapeinside=||]{C++}
BCL::HashMap<int, int> map(size);

BCL::HashMapBuffer<int, int> buffer(map, queue_size,
                                    message_size);

for (...) {
 buffer.insert(key, value);
}

buffer.flush();

\end{minted}
\vspace{-0.5em}
  \caption{A small change to user code---inserting into the \texttt{HashMapBuffer}
  instead of the \texttt{HashMap}---causes inserts to be batched together.}
  \label{fig:hashbuff}
  \vspace{-1em}
\end{figure}

\subsubsection{Blocked Bloom Filter}
\label{sec:blockedbloom}
Instead of being comprised of a single bit array, \textit{blocked Bloom filters}
are composed of many smaller Bloom filters \cite{putze2009cache}.  To insert a
value into a blocked
Bloom filter, the value is first hashed to determine which Bloom filter the
value should be stored in.  The item will then be hashed $k$ times to determine
which bits in the smaller Bloom filter need to be set.  In shared memory systems,
blocked Bloom filters are used to improve the cache performance of large Bloom filters
with a block size that is a multiple of the cache line size.

In \bcl{}, we implement a distributed blocked Bloom filter as
\texttt{BCL::BloomFilter} to solve both of the
issues with distributed Bloom filters raised above.  Our distributed blocked
Bloom filter consists of a number of 64-bit Bloom filters.  Inserting an item
into our blocked Bloom filter now requires, in all cases, a single atomic
memory operation, and the operation is fully atomic.
Checking if an operation is present in the blocked Bloom filter requires one
remote read memory operation.


\section{\bcl{} \container{}s}
\label{sec:container}
All \bcl{} data structures use \bcl{} \container{}s, which provide a transparent
abstraction for storing complex data types in distributed memory with low overhead.
\bcl{} \container{}s are necessary because not all data types can be stored in
distributed memory by byte copying.  The common case for this is a struct or
class, such as the C++ standard library's \texttt{std::string}, which contains
a pointer.  The pointer contained inside the class is no longer meaningful once
transferred to another node, since it refers to local memory that is now
inaccessible, so we must use some other method to serialize and deserialize our
object in a way that is meaningful to remote processes.
At the same time, we would like to optimize for the common
case where objects \textit{can} be byte copied and avoid making unnecessary
copies.

\subsubsection{Implementation}
\bcl{} \container{}s are implemented using the C++ type system.  A \bcl{}
\container{} is a C++ struct that takes template parameters \texttt{T}, a
type of object that the \container{} will hold, and \texttt{TSerialize},
a C++ struct with methods to serialize objects of type \texttt{T}
and deserialize stored objects back to type \texttt{T}.  \bcl{} \container{}s
themselves are of a fixed size and can be byte copied to and from shared memory.
An \container{} has a \textit{set} method, which allows the user to store an
object in the \container{}, and a \textit{get} method, which allows the user
to retrieve the object from the container.

\bcl{} includes a number of \texttt{TSerialize} structs for common C++ types,
and these will be automatically detected and utilized by the \bcl{} data
structures at runtime.
Users will usually not have to write their own serialization and
deserialization methods unless they wish to use custom types which use heap
memory or other local~resources.

A finer point of \bcl{} serialization structs is that they may serialize objects
to either \textit{fixed length} or \textit{variable length} types.  This is
handled automatically at compile time by looking at the return type of the
serialization struct: if the serialization struct returns an object of any
normal type, then the serialized object is taken to be fixed size and is stored
directly as a member variable of the serialization struct.  If, however, the
serialization struct returns an object of the special type
\texttt{\bcl{}::serial\_ptr}, this signifies that the object is
\textit{variable length}, even when serialized, so we must instead store a
global pointer to the serialized object inside the \container{}.

\subsubsection{User-Defined Types}
To store user-defined types in \bcl{} data structures, users can simply define
serialization structs for their type and inject the struct into the \bcl{}
namespace.  For byte-copyable types, this struct can be an empty struct that
inherits from the \texttt{\bcl{}::identity\_serialize <T>} template struct.

\subsubsection{Copy Elision Optimization}
\label{sec:copy_elision}
An important consideration when using serialization is overhead in the common
case, when no serialization is actually required.  In the common byte-copyable
case, where the serialization struct simply returns a reference to the original
object, intelligent compilers are able to offer some \textit{implicit} copy elision
automatically.  We have observed, by examining the assembly produced, that the GNU
and Clang compilers are able to optimize away the unnecessary copy when a
\container{} object is retrieved from distributed memory, \texttt{get()} is
called to retrieve the item lying inside.  However, when an array of items is
pulled together from distributed memory and unpacked, the necessary loop
complicates analysis and prevents the compiler from performing this copy elision.

For this reason, \bcl{} data structures perform \textit{explicit} copy elision
when reading or writing from an array of \container{}s stored in distributed memory
when the \container{} inherits from the \texttt{\bcl{}::identity\_serialize <T>} struct,
which signifies that it is byte copyable.  This is a compile-time check, so
there is no runtime cost for this optimization.

\section{Concurrency Promises}
\label{sec:conprom}
As we have shown for various \bcl{} data structures,
distributed data structure operations often have multiple alternate implementations,
only some of which will be correct in the context in which an operation is issued.
A common example of this is \textit{phasal} operations, where data structures
are manipulated in separate phases, each of which is separated by a barrier.
Figures \ref{fig:sortex} and \ref{fig:hashbuff} both demonstrate phasal operations.
Crucially, the barriers associated with phasal operations provide atomicity
between different types of operations that often allows the use of implementations
with fewer atomicity guarantees.  For example, a find operation in a hash
table can be executed with a more optimized implementation---a single remote get
operation, rather than 2 AMOs and a remote get---when we are guaranteed that only
find operations will be executed in the same barrier region.

To allow users to take advantage of these optimized implementations in a
straightforward manner, we allow users to optionally provide
\textit{concurrency promises}, which are lists of data structure operations
that could take place concurrently with the operation being issued.  To use an
optimized version of hash table \texttt{find}, we can pass as an extra argument
to the \texttt{find} function the value \texttt{ConProm::HashMap::find}.  This
indicates that only \texttt{find} operations may occur simultaneously with this
invocation.  Similarly, if in a particular context a \texttt{find} operation
might also occur concurrently with a \texttt{insert} operation, we can pass the
concurrency promise \texttt{ConProm::HashMap::find | ConProm::HashMap::insert}.

It's important to note that, since C++ template metaprogramming does not have
full-program knowledge (unless the whole program is expressed as a single
expression), it is not possible to automatically identify these optimization
opportunities and pick correct implementations with a library.  Instead, it would
require static analysis using either a preprocessing tool or a separate parallel
programming language with an intermediate representation that preserves semantic
knowledge of data structure operations.  Our approach here is to provide a
facility for the user to easily specify invariants, rather than to identify
them automatically.

\section{Backends}
\bcl{} backends implement a limited number of communication primitives to
provide full support for the \bcl{} Core.  These include an \texttt{init()}
function which allocates symmetric shared memory segments of a requested size,
the \texttt{barrier()},
\texttt{read()} and \texttt{write()} operations that perform variable-sized
reads and writes to global memory, at least an atomic compare-and-swap, and
broadcast and reduce operations.

\begin{table}
  \centering
  \begin{tabular}{ l | l | l }
    Name & Processors & Interconnect\\
    \hline
    Cori Phase I & Intel Xeon Haswell & Cray Aries\\
    Summitdev & IBM POWER8 & Mellanox EDR Infiniband\\
    AWS c3.4xlarge & Intel Xeon & 10Gb/s Ethernet\\
  \end{tabular}
    \vspace{1em}
  \caption{\textsc{Summary of systems used in evaluation.}}
      \vspace{-2em}
  \label{table:systems}
\end{table}

\section{Experimental Evaluation}
\label{sec:eval}
We evaluated the performance of \bcl{}'s data structures using ISx, an integer
sorting mini-application, Meraculous, a mini-application taken from large-scale
genome
assembly, and a collection of microbenchmarks.  In order to evaluate the
performance portability of \bcl{} programs,
we tested the first two benchmarks across three different computer systems, as
outlined
in Table \ref{table:systems}.  On Cori, experiments are performed up to 512 nodes.
On Summitdev, experiments are performed up to 54 nodes, which is the size of the
whole cluster.  On AWS, we provisioned a 64 node cluster and performed scaling
experiments up to its full size.  For reasons of space, the microbenchmarks
are presented only on Cori up to 64 nodes.

\begin{figure*}
  \centering
  \begin{tabular}{c c c}
    \includegraphics[width=0.31\linewidth]{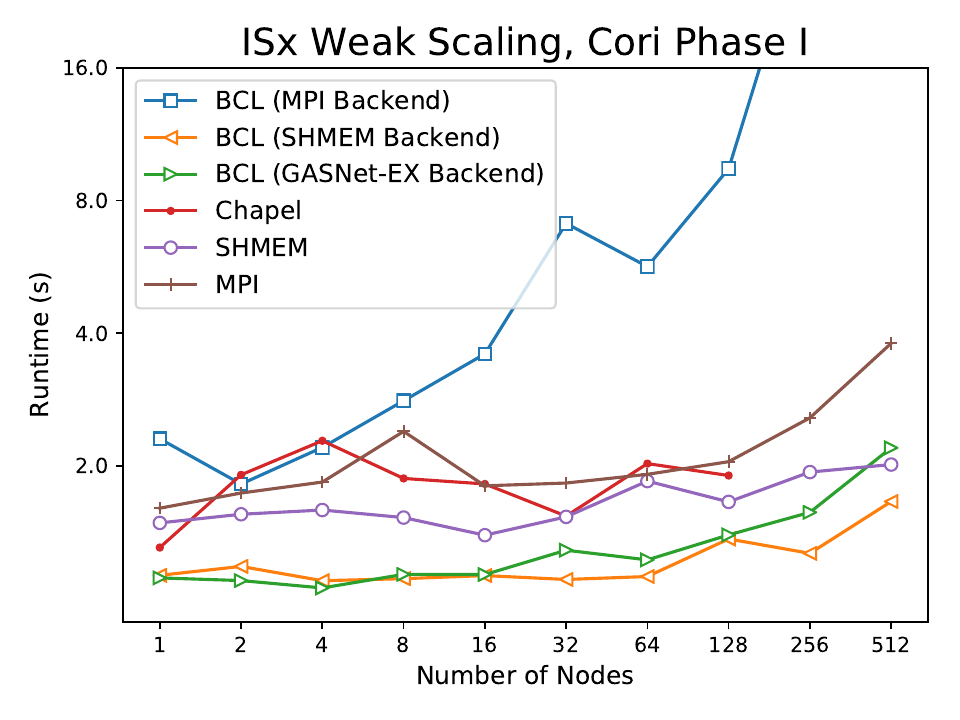} &
    \includegraphics[width=0.31\linewidth]{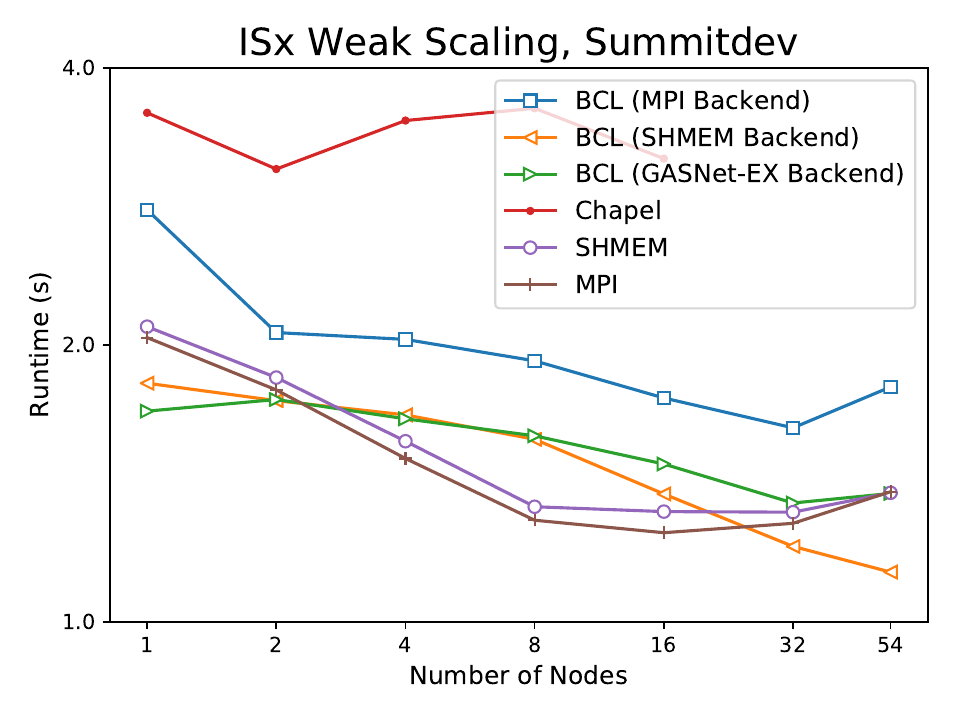} &
    \includegraphics[width=0.31\linewidth]{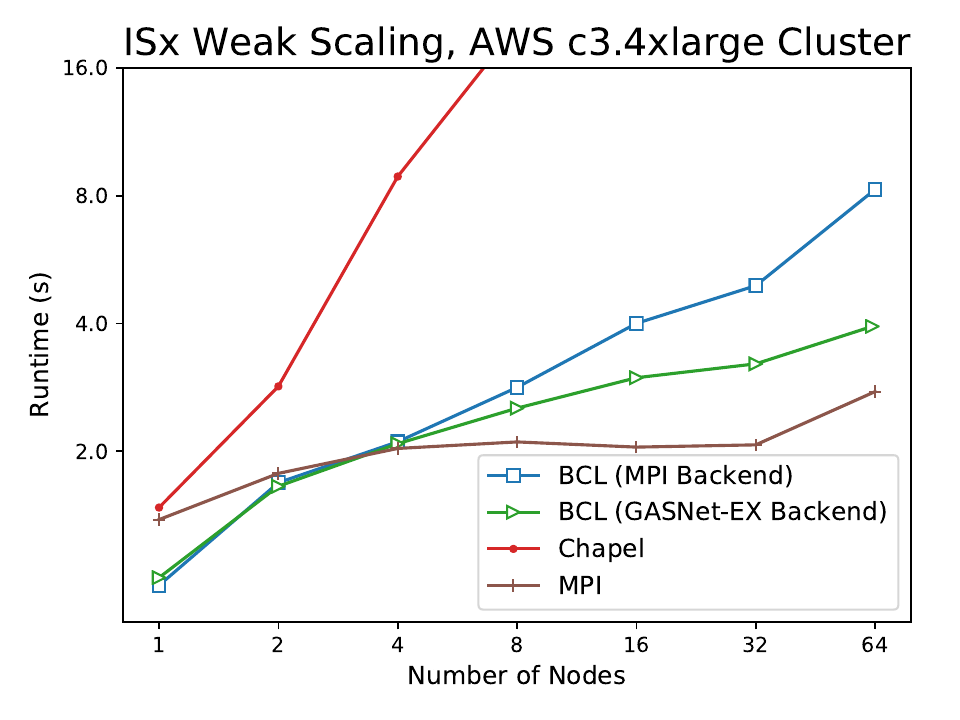}\\
  \end{tabular}
  \vspace{-1em}
  \caption{Performance comparison on the ISx benchmark on three different
  computing systems.  All runs measure weak scaling with $2^{24}$ items per process.}
  \label{fig:isx-sort}
\end{figure*}

\subsection{ISx Benchmark}
\label{sec:cqeval}
To test out our queue's performance, we implemented a
bucket sort algorithm to execute the ISx benchmark \cite{hanebutte2015isx}.
The ISx benchmark is a simple bucket sort benchmark performed on
uniformly distributed data.  It consists of two stages, a distribution stage and
a local sort stage.  In the distribution stage, processes use pre-existing
knowledge about the distribution of the randomly generated data to assign each
key to be sorted to a bucket, where there is by default one bucket on each node.
After this stage, each process simply performs a local sort on its received data.
The original ISx benchmark comes with an MPI implementation, which uses an
all-to-all collective for the distribution stage and an OpenSHMEM implementation,
which sends data asynchronously.  An implementation in Chapel, a high-level
parallel programming language, has also been published~\cite{hemstad2016study,
chapelisx}.

\subsubsection{\bcl{} Implementation}
We implemented our bucket sort in \bcl{} using the circular queue data structure.
First, during initialization, we place one circular queue on each process.  During
the distribution phase, each process pushes its keys into the appropriate remote
queues.  After a global barrier, each node performs a local sort on the items
in its queue.  During the
distribution phase, we perform \textit{aggregation} of inserts to amortize the
latency costs of individual inserts.  Instead of directly pushing individual items
to the remote queues, we first place items in local buffers corresponding to
the appropriate remote queue.  Once a bucket reaches a set message size, we
push the whole bucket of items at once and clear the local bucket.  It's important
to note that this push is \textit{asynchronous}, meaning that the communication
involved with pushing items to the queue can be overlapped with computation
involved with sorting the items.
The fact
that \bcl{} circular queue's push method accepts a vector of items to insert
simultaneously makes adding aggregation very straightforward.  Even with this
optimization, our full \bcl{} sorting benchmark code, including initialization
and timing, is only 72 lines long, compared to the original MPI and SHMEM
reference implementations at 838 and 899 lines, and the Chapel implementation at
244 lines.  A slightly abbreviated version of our implementation is listed in
Figure \ref{fig:sortex}.

\subsubsection{Analysis}
As shown in Figure \ref{fig:isx-sort}, our \bcl{} implementation of ISx performs
competitively with the reference and Chapel~implementations.

On the Cray Aries systems, \bcl{} outperforms the other implementations.  This is
because \bcl{} is able to overlap communication with computation: the asynchronous
queue insertions overlap with sorting the values into buckets.  This is an
optimization that would be much more complicated to apply in a low-level MPI or
SHMEM implementation (the reference implementation uses all-to-all patterns),
but is straightforward using \bcl{}'s high-level interface.

There is an upward trend in the \bcl{} scaling curves toward the high extreme
of the graph on Cori.
This is because as the number of processes increases, the number of values sent to
each process decreases.  At 512 nodes with 32 processes per node, each process
will send, on average, 1024 values to each other process.  With our message size
of 1024, on average only one push is sent to each other process, and the potential
for communication and computation overlap is much smaller, thus our solution
degenerates to the synchronous all-to-all solution, and our performance matches
the reference SHMEM implementation.  Note that performance with the MPI backend is poor
on Cori; we believe this is because the MPI implementation is failing to use
hardware atomics.

The performance on Summitdev is similar, except that there is a
slight downward trend in all the scaling lines because of cache effects.
As the number of processes increases, the keyspace on each node decreases, and the
local sort becomes more cache efficient.

Historically, PGAS programs have not faired well on Ethernet clusters, since PGAS
programs often rely on fast hardware RDMA support.  With our \bcl{} implementation,
we are able to increase the message size to amortize the cost of slow atomic
operations.  While our performance on AWS does not scale as well as the reference MPI
implementation, we consider the performance acceptable given that it is a
high-level implementation running in an environment traditionally deemed the
exclusive domain of message-passing.  On the Ethernet network, the GASNet-EX
backend, which is using the UDP conduit, performs better than the MPI backend,
which is using Open MPI.

\begin{figure*}
  \begin{tabular}{c c c}
    \includegraphics[width=0.31\linewidth]{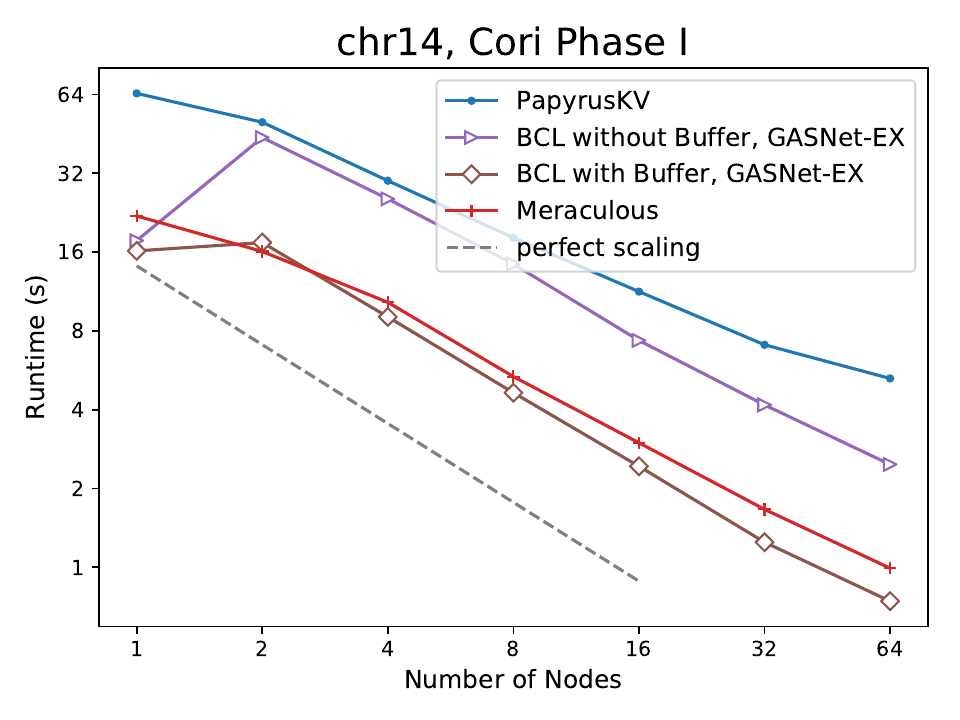} &
    \includegraphics[width=0.31\linewidth]{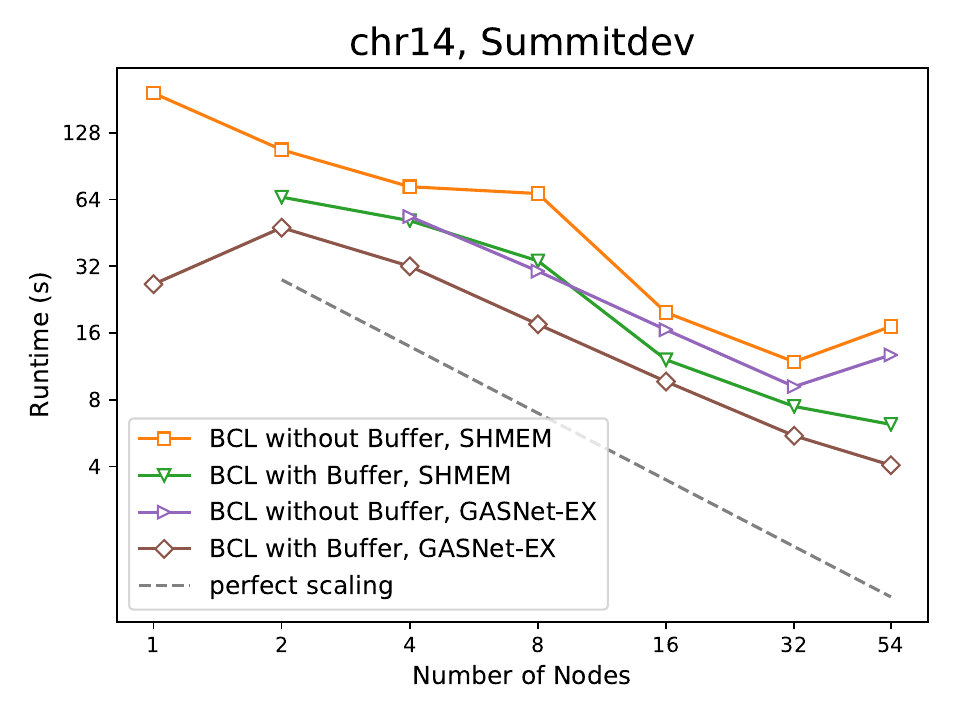} &
    \includegraphics[width=0.31\linewidth]{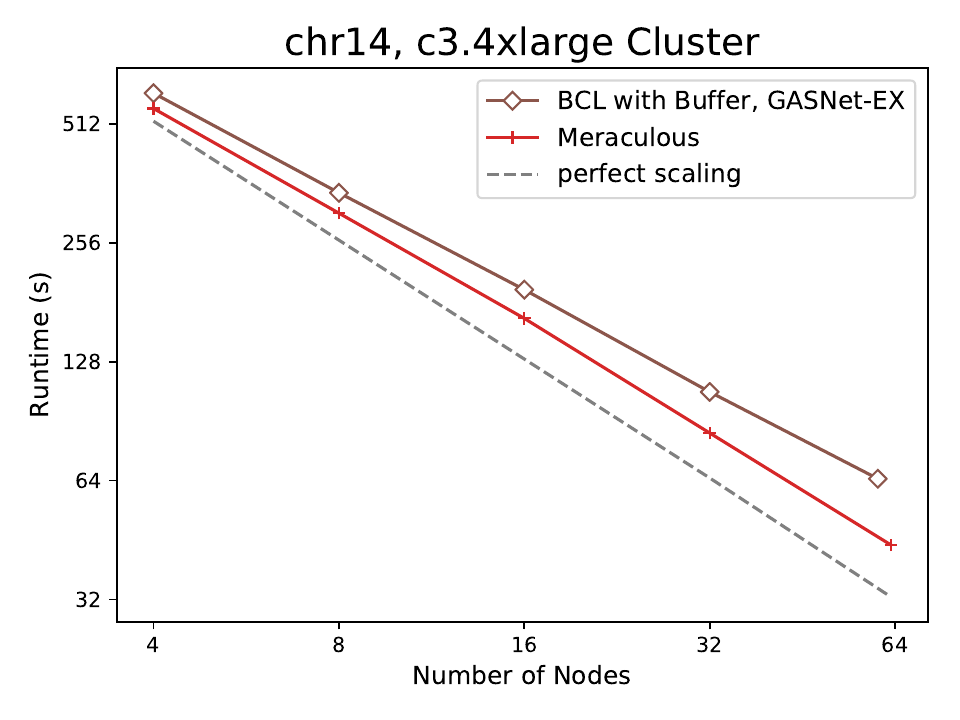}\\
  \end{tabular}
    \vspace{-1em}
  \caption{Performance comparison on the Meraculous benchmark on the
           \textit{chr14} dataset.}
  \label{fig:comparison}
\end{figure*}

\begin{figure}
  \centering
  \includegraphics[width=0.875 \linewidth]{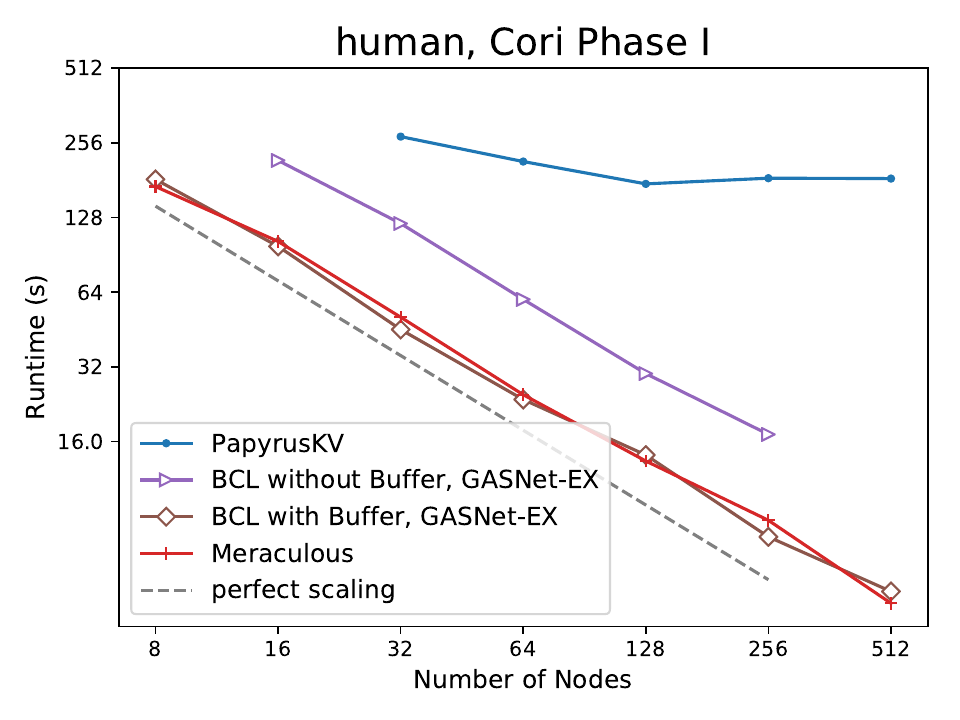}
    \vspace{-1em}
  \caption{Performance on the Meraculous benchmark}
  \vspace{-1em}
  \label{fig:human}
\end{figure}

\subsection{Genome Assembly}
We evaluated \bcl{}'s generic hash table using two benchmarks taken from a
large-scale scientific application, a de novo genome assembly pipeline: contig
generation and $k$-mer counting.  Both use a
hash table, contig generation to traverse a de Bruijn graph of overlapping
symbols, and $k$-mer counting to compute a histogram describing the number of
occurrences of each $k$-mer across reads of a DNA sequence.
\subsubsection{Contig Generation in De Novo Genome Assembly}
During the \textit{contig generation} stage of de novo genome assembly,
the many error-prone reads recorded by a DNA sequencer have been condensed into
$k$-mers, which are short error-free strands of DNA guaranteed to overlap each
other by exactly $k$ bases.  The goal of contig generation is to process these
$k$-mers to produce \textit{contigs}, which are long strands of contiguous
DNA~\cite{chapman2011meraculous}.

Assembling these $k$-mers into longer strands of DNA involves using a hash
table to traverse the de Bruijn graph of overlapping $k$-mers.  This is performed
by taking a $k$-mer, computing the next overlapping $k$-mer in the sequence, and
then looking it up in the hash table.  This process is repeated recursively
until a $k$-mer is found which does not match the preceding $k$-mer or a $k$-mer
with an invalid base is discovered.

A fast implementation for contig generation is relatively simple in a serial program,
since using any of a large number of generic hash table libraries will yield high
performance.
However, things are not so simple in distributed memory.  The reference
solution for Meraculous, written in UPC, is nearly 4,000 lines long
and includes a large amount of boilerplate
C code for operations like reading and writing to memory buffers \cite{nerscmeraculous}.

\label{sec:hasheval}

The implementation of the contig generation phase of a genome assembly pipeline
is greatly simplified by the
availability of a generic distributed hash table in \bcl{}.  As described above,
the contig generation benchmark is really a simple application split into two
phases, an insert phase, which builds the hash table, and a traversal phase,
which uses the hash table to traverse the de Bruijn graph of overlapping symbols.
Because of this phasal behavior, we are able to optimize the performance of the
hash table using \bcl{}'s hash map buffer, which groups together inserts by
inserting them all at once into a local queue on the node where they will likely
be placed, then inserting them all using a fast local insert when a flush operation
is called on the hash map buffer.  Our implementation of the Meraculous benchmark
is only 600 lines long, 400 of which consist of code for reading, parsing, and
manipulating $k$-mer objects.

We implemented the contig generation phase of a genome assembly pipeline using
the Meraculous algorithm \cite{georganas2014parallel,georganas2015hipmer,
chapman2011meraculous}.  Our implementation is similar to the high-performance
UPC implementation \cite{georganas2014parallel}, but (1) uses our
generic hash table, instead of a highly specialized hash table, and (2) uses a
less sophisticated locking scheme, so sometimes processes may redundantly
perform extra work by reconstructing an already constructed contig.
We should note that the Meraculous UPC benchmark is based on the HipMer application, which
may have higher performance \cite{georganas2015hipmer}.

We benchmarked our hash table across the same three HPC systems described in
Table
\ref{table:systems} using the \textit{chr14} dataset, which is
from sequencing data of human chromosome 14.
We compared our implementation to
the high-performance UPC reference Meraculous benchmark implementation provided
on the NERSC website, which we compiled with Berkeley UPC with hardware atomics
enabled~\cite{nerscmeraculous,georganas2014parallel}.
We also compared our hash table to PapyrusKV, a high-performance general-purpose
hash table implemented in MPI which has a Meraculous benchmark implementation
available \cite{kim2017papyruskv}.  All performance results were obtained by
running one process per
core.  Benchmarks for the UPC implementation are not available on Summitdev because
the code fails on POWER processors due to an endianness issue.
We also used the Meraculous benchmark prepared by the developers
of PapyrusKV \cite{kim2017papyruskv}.  As shown in Figure \ref{fig:comparison},
our \bcl{} implementation of the contig generation
benchmark matches or exceeds the performance of both the reference high-performance
implementation of Meraculous and the general PapyrusKV hash table.
In Figure \ref{fig:human}, we also show a larger scaling experiment using the
human genome as a dataset on Cori.  Our \bcl{} implementation appears to be
both scalable to high numbers of nodes and portable across different architectures
and~interconnects.

\begin{figure}
  \centering
  \includegraphics[width=0.875 \linewidth]{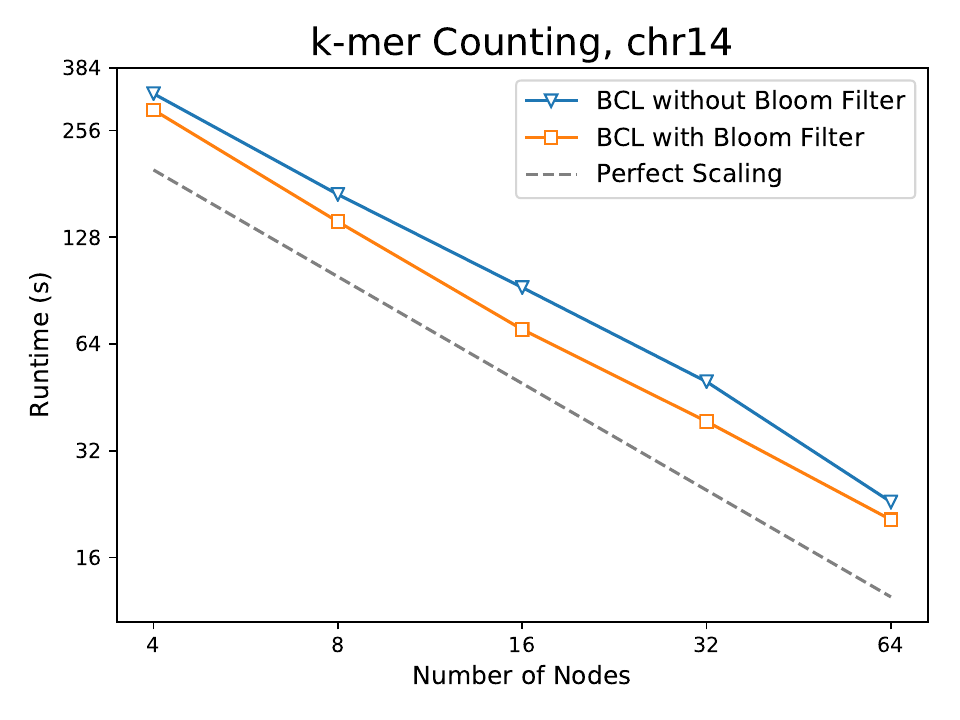}
    \vspace{-1em}
  \caption{Strong scaling for our $k$-mer counting benchmark using dataset
  \textit{chr14}.}
    \vspace{-1em}
  \label{fig:countperf}
\end{figure}

\subsubsection{$k$-mer Counting}
$k$-mer counting is another benchmark from de novo genome assembly.  In $k$-mer
counting, we examine a large number of lossy reads of DNA sequences and split
these reads up into small, overlapping chunks of length $k$.  We then use a hash
table to calculate a histogram, accumulating the number of occurrences of each
individual $k$-mer, to try to eliminate errors.  A large number
of $k$-mers will be erroneous, and, as a result, will appear only once
in the counting.

To speed up this histogram calculation,
we can avoid unnecessary hash table
lookups for items which appear only once in the hash table by using our
distributed blocked Bloom filter as discussed in Section \ref{sec:blockedbloom}.
With this optimization, we first atomically insert a $k$-mer into the Bloom filter,
then only update its histogram value if the $k$-mer was already present in the
Bloom filter. This optimization also significantly reduces the overall memory consumption
of k-mer counting because
a high portion of the unique $k$-mers occur only once due to sequencing errors.
Consequently, Bloom filters are now common in single node $k$-mer counters~\cite{melsted2011efficient}.
However, it is harder to efficiently take advantage of Bloom filters in distributed $k$-mer counting. In the
absence of an efficient distributed Bloom filter that keeps global information about the $k$-mers processed so far,
all the occurrences of a $k$-mer had to be localized in and counted by the same process for
local Bloom filters to produce accurate counts~\cite{georganas2014parallel}. \bcl{}'s distributed Bloom
filter avoids this localization and the expensive all-to-all exchange of all $k$-mers associated with~it.

As shown in Figure \ref{fig:countperf}, our $k$-mer counting benchmark shows
excellent strong scaling, along with a slight performance boost when using our
distributed blocked Bloom filter. Aside from this performance boost, lower memory footprint
is another benefit of \bcl{}'s distributed Bloom filter, since $k$-mers which
appear only once need not be inserted into the hash table.

\begin{figure}
  \centering
  \begin{tabular}{c c}
    \includegraphics[width=0.48\linewidth]{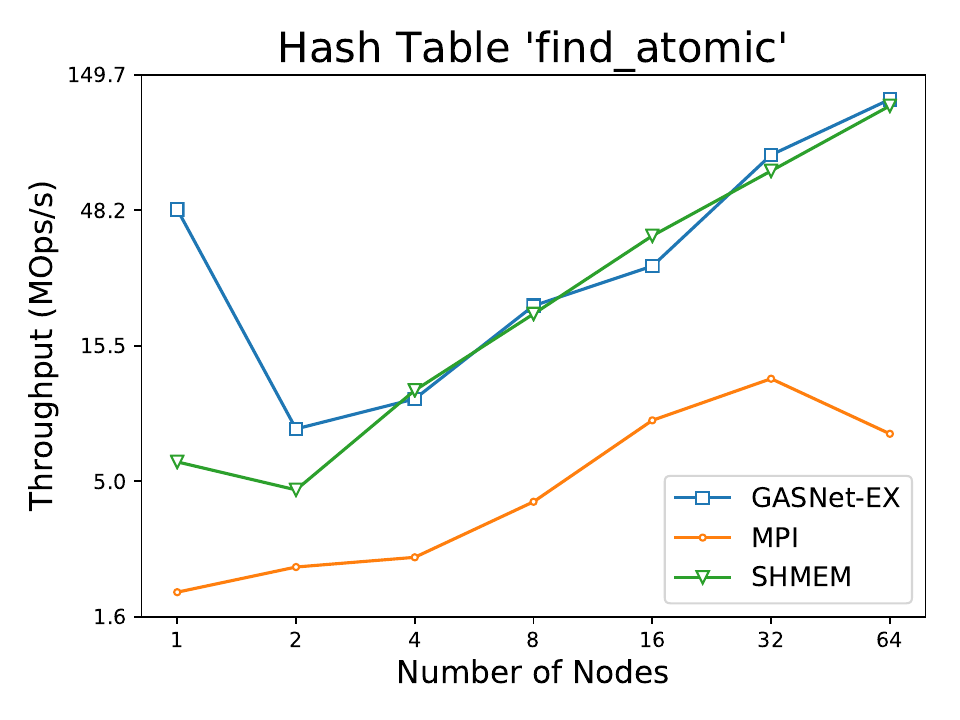} &
    \includegraphics[width=0.48\linewidth]{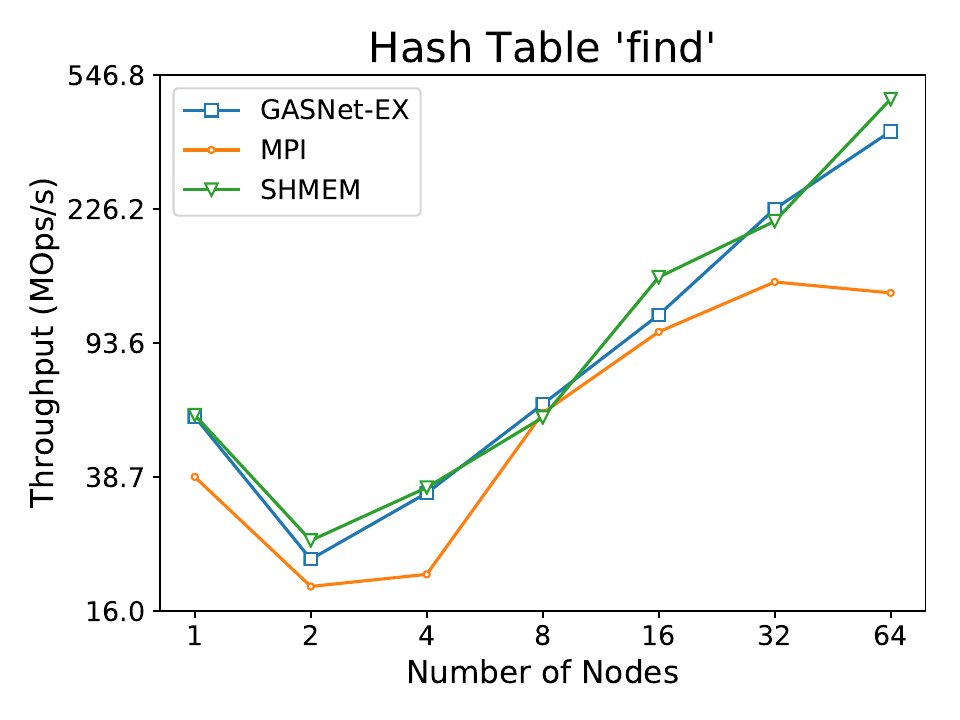}\\
    \includegraphics[width=0.48\linewidth]{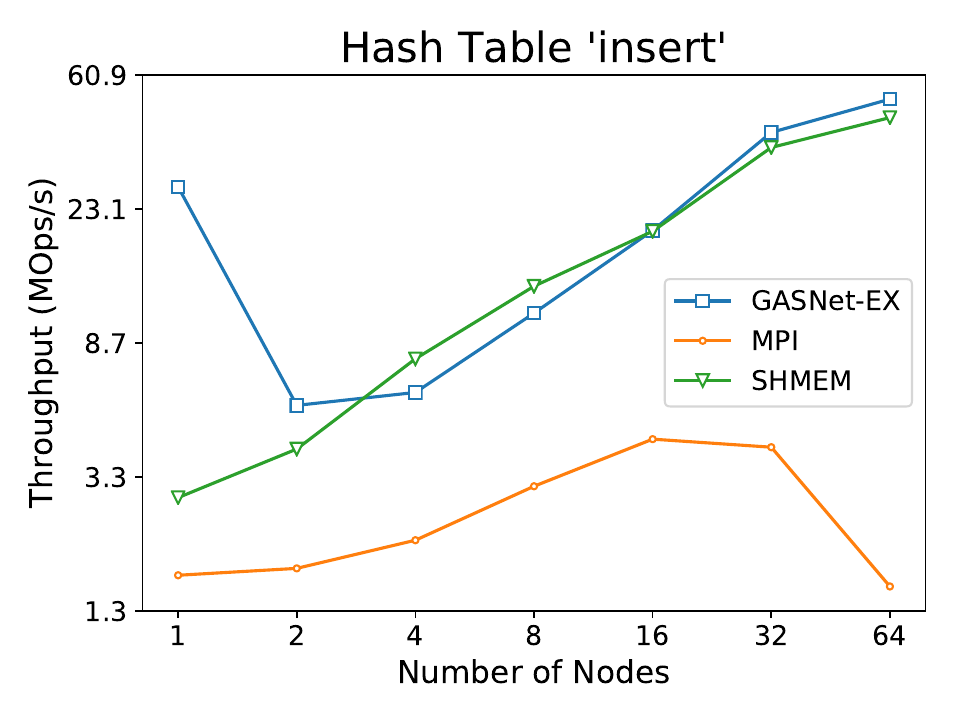} &
    \includegraphics[width=0.48\linewidth]{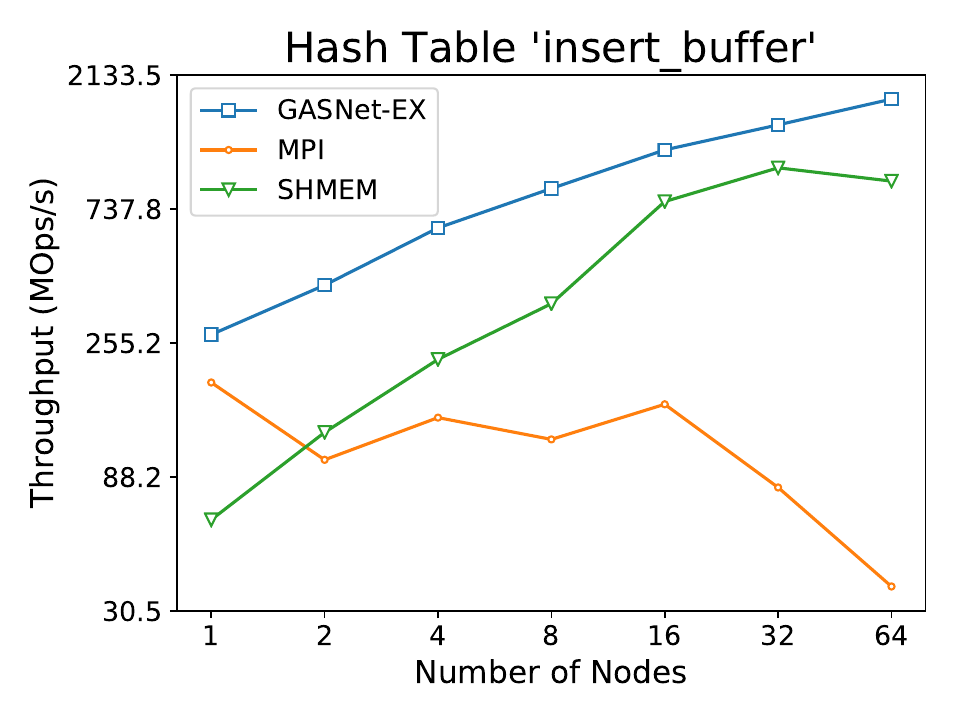}\\
  \end{tabular}
    \vspace{-1em}
  \caption{Microbenchmarks for the hash table.}
  \label{fig:hashbench}
\end{figure}

\begin{figure}
  \centering
  \begin{tabular}{c c}
    \includegraphics[width=0.48\linewidth]{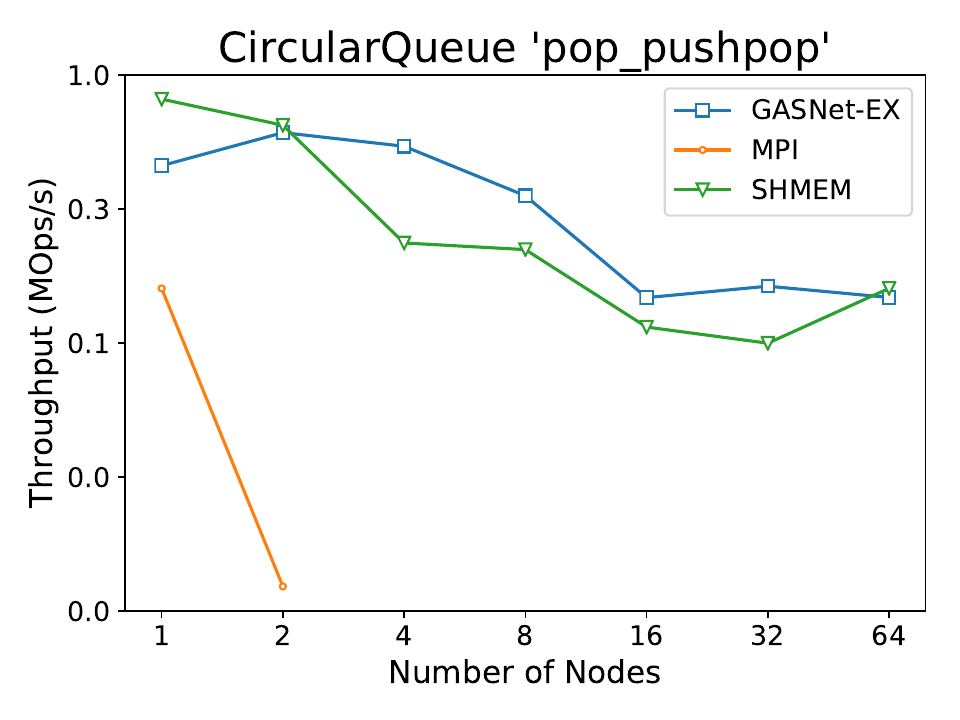} &
    \includegraphics[width=0.48\linewidth]{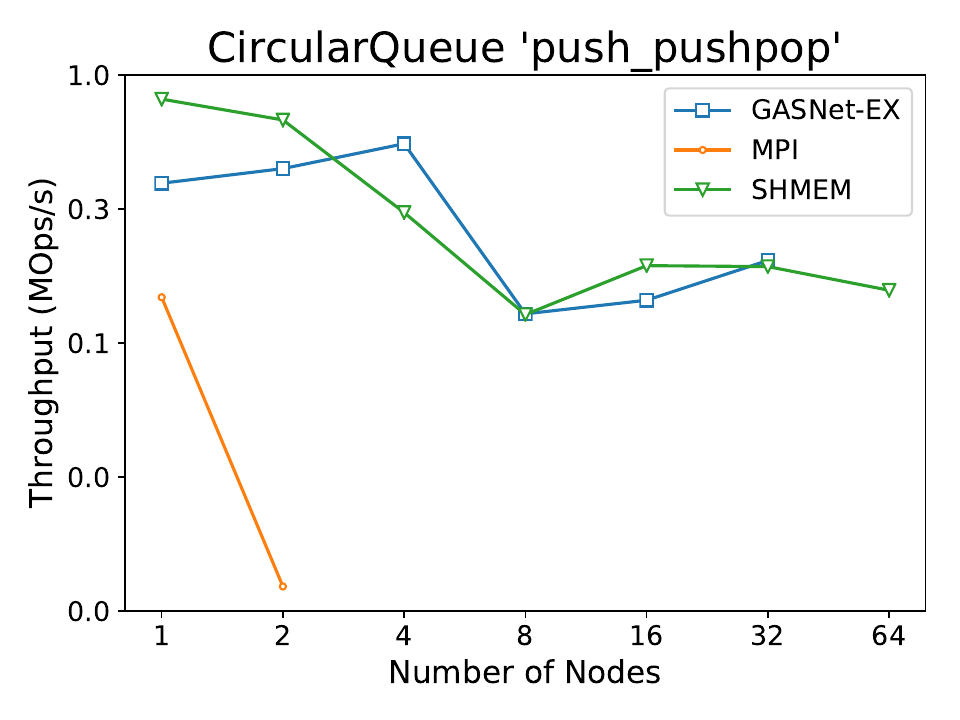} \\
    \includegraphics[width=0.48\linewidth]{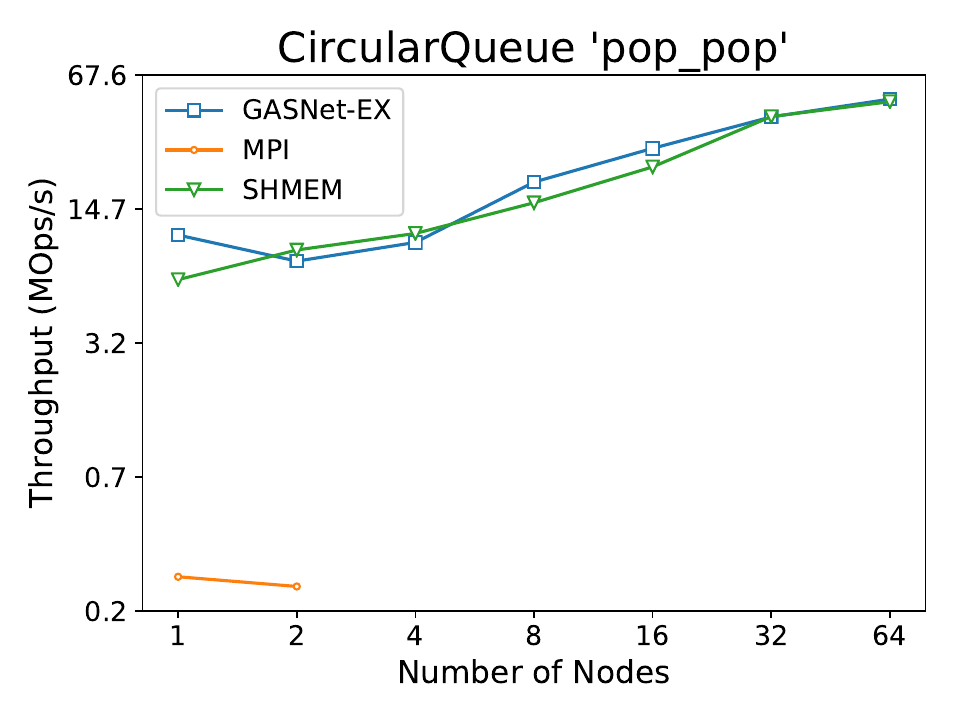} &
    \includegraphics[width=0.48\linewidth]{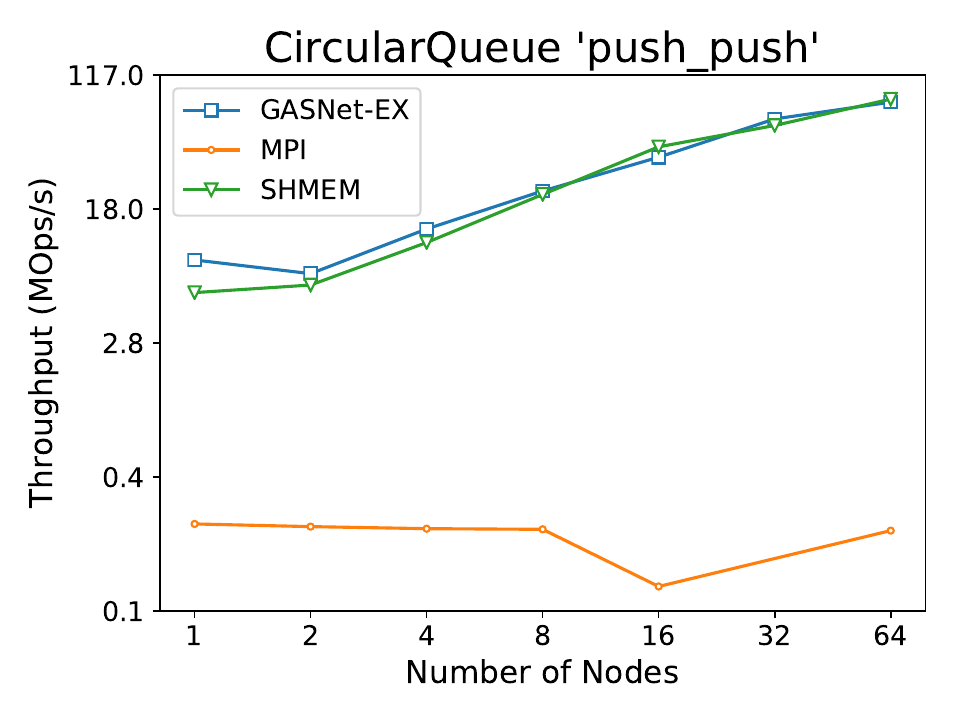} \\
    \includegraphics[width=0.48\linewidth]{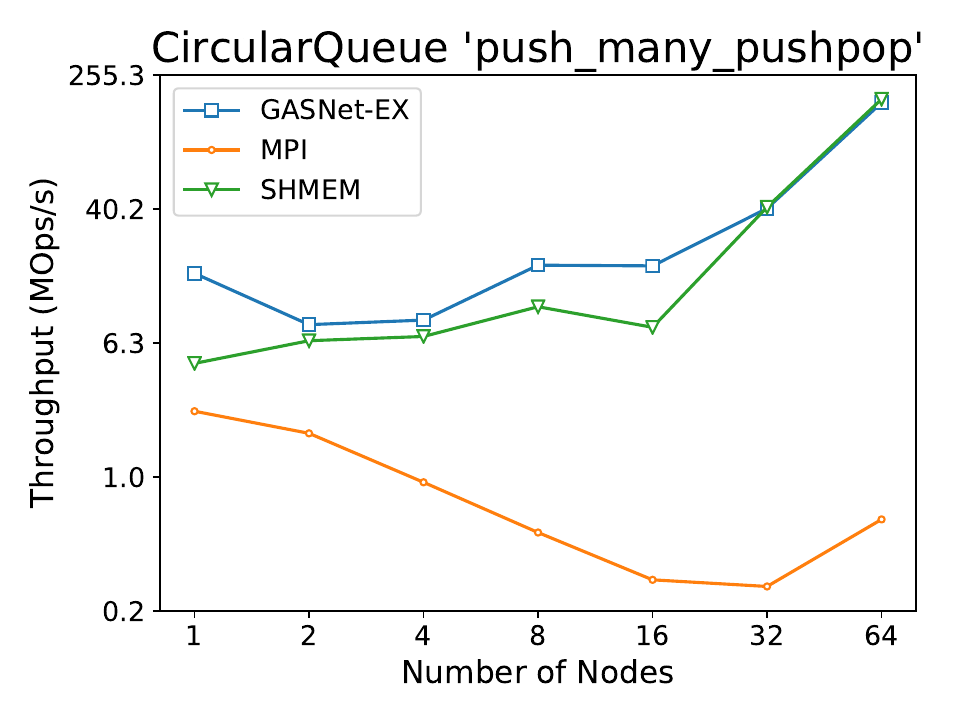} &
    \includegraphics[width=0.48\linewidth]{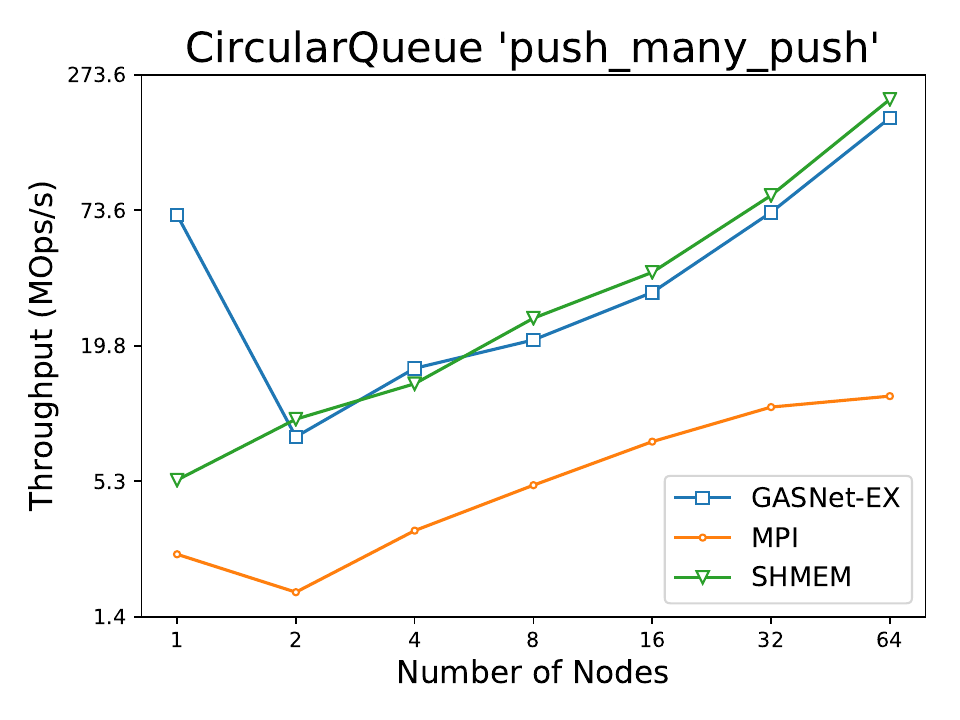} \\
  \end{tabular}
    \vspace{-1em}
  \caption{Microbenchmarks for CircularQueue.}
  \label{fig:circbench}
\end{figure}

\begin{figure}
  \centering
  \begin{tabular}{c c}
    \includegraphics[width=0.48\linewidth]{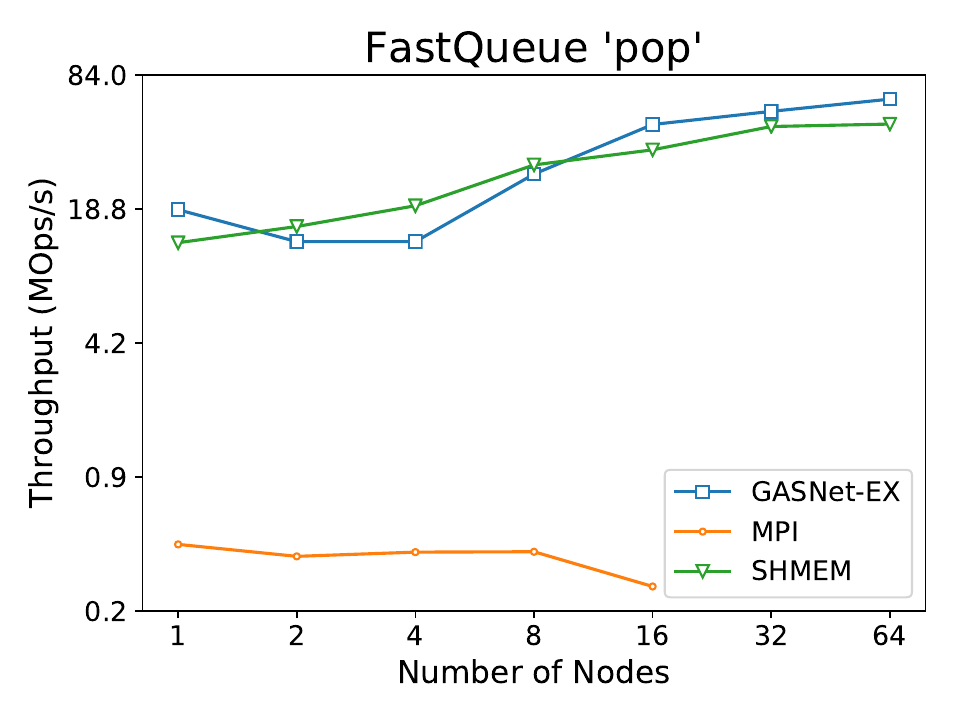} &
    \includegraphics[width=0.48\linewidth]{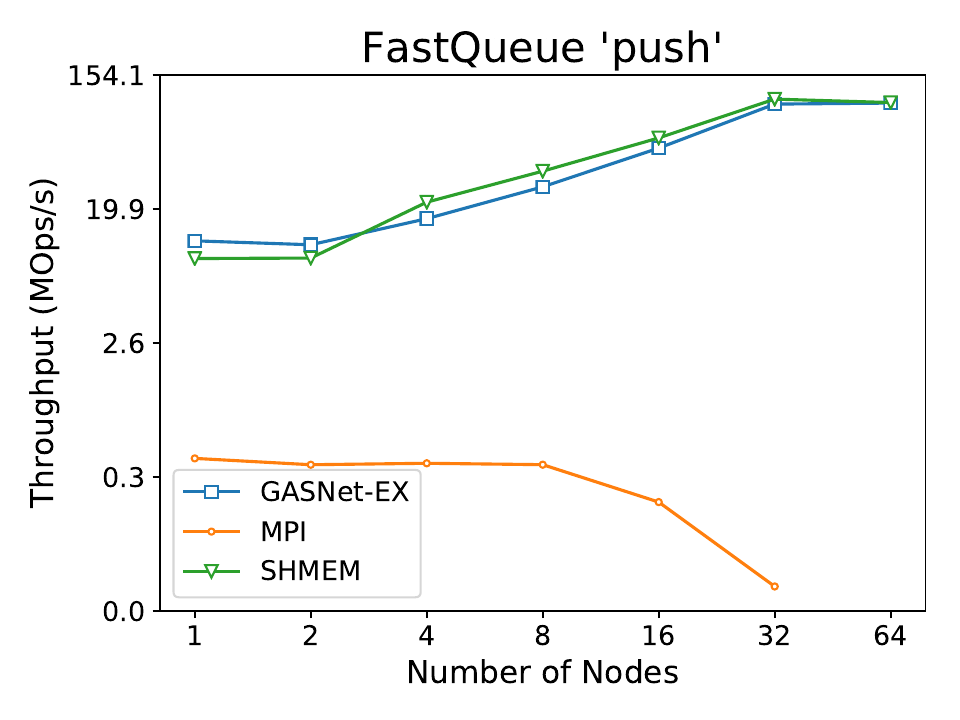} \\
    \includegraphics[width=0.48\linewidth]{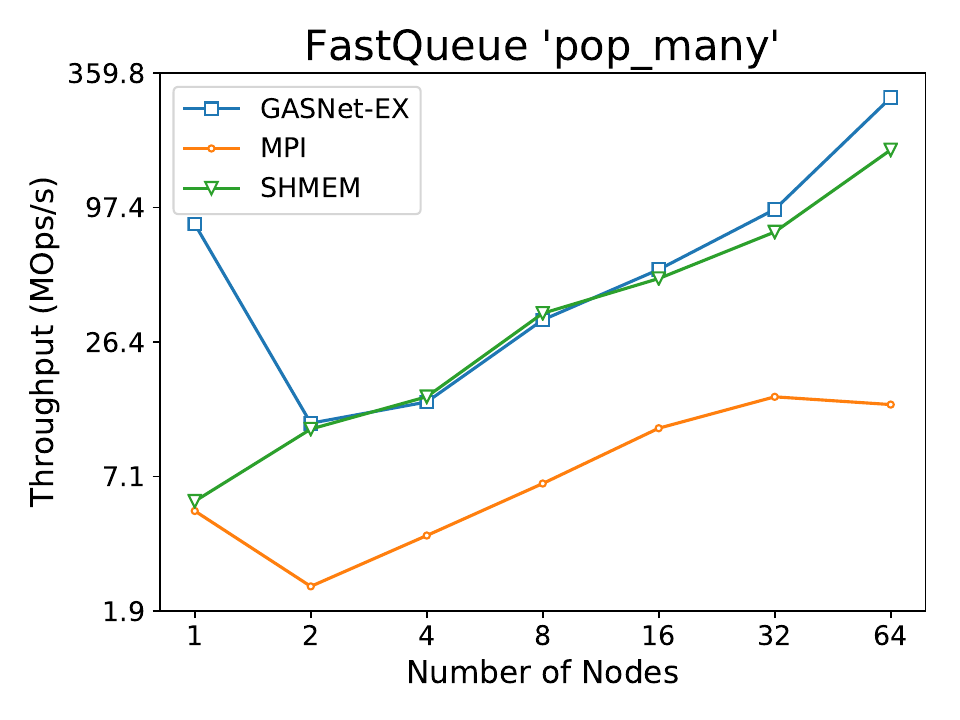} &
    \includegraphics[width=0.48\linewidth]{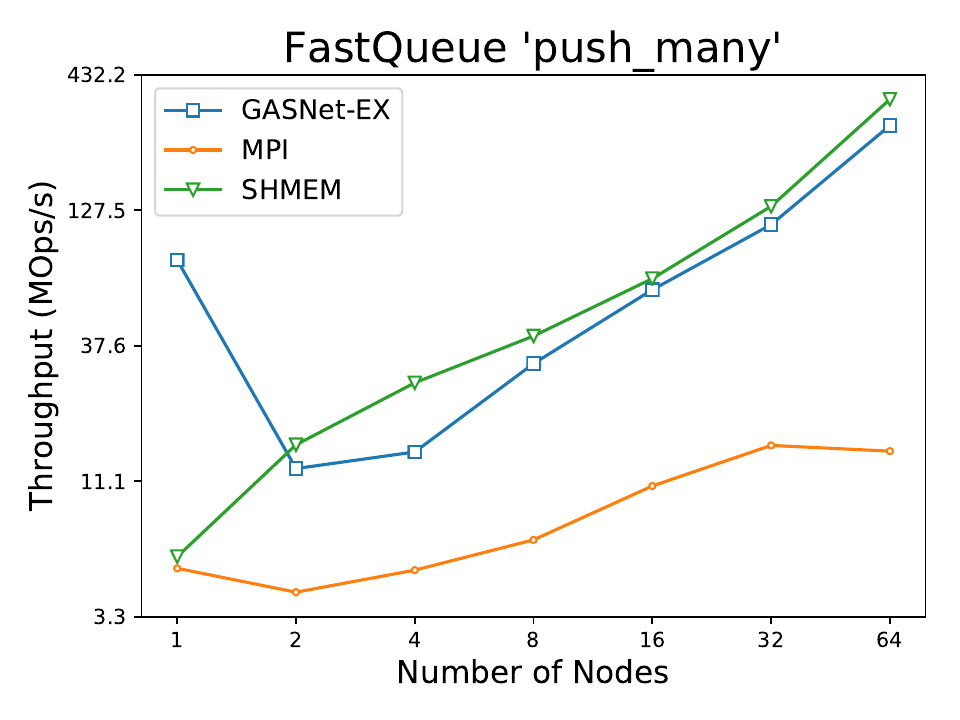} \\
  \end{tabular}
    \vspace{-1em}
  \caption{Microbenchmarks for FastQueue.}
  \label{fig:fastbench}
\end{figure}

\subsection{Microbenchmarks}

We prepared a collection of microbenchmarks to compare (1) different backends'
performance across data structure operations and (2) the relative performance of
different implementations of data structure operations.  Each benchmark tests a
single data structure operation.  Turning first to the
\texttt{HashMap} microbenchmarks in Figure \ref{fig:hashbench}: we see clear differences between fully atomic
versions of data structure operations (\texttt{find\_atomic} and \texttt{insert})
and versions offering fewer atomicity guarantees or buffering (\texttt{find} and
\texttt{insert\_buffer}).  We see that buffering offers an order of magnitude
increase in performance, which we would expect from transforming a latency-bound
operation into a bandwidth-bound operation, while the optimized \texttt{find}
operation offers a factor of 2-3x improved performance, as we would expect from
the relative best-case costs ($2A + R$ and $R$).

The queue performance features two kinds of benchmarks: benchmarks looking at
operations by all processes on a single queue, and benchmarks looking at
operations by all processes on a collection of queues, one on each processor
(the latter benchmarks are labeled ``many'').  In the \texttt{CircularQueue}
benchmarks, we see that fully atomic operations (\texttt{pop\_pushpop} and \texttt{push\_pushpop})
are quite expensive when all processes are inserting into a single queue, compared
to the less-atomic \texttt{pop\_pop} and \texttt{push\_push}.  This is unsurprising,
since the final compare-and-swap operation in an atomic \texttt{CircularQueue}
push or pop essentially serializes the completion of operations.  When pushing
to multiple queues, using the less-atomic operation gives a factor of two performance
improvement (\texttt{push\_many\_pushpop} vs. \texttt{push\_many\_push}).

Looking at the \texttt{FastQueue} benchmarks, we see that this data structure
does achieve a significant performance improvement even over the less-atomic
implementations of data structure operations in \texttt{CircularQueue}.

Across all benchmarks, it appears that GASNet-EX is most effective at automatically
utilizing local atomics when only running on one node, while MPI lags behind on
most benchmarks, particularly those which make heavy use of atomic operations.

\section{Related Work}
UPC, Titanium, X10, and Chapel are parallel programming languages which offer
a PGAS distributed memory abstraction~\cite{yelick1998titanium,upc2005upc,
charles2005x10,weiland2007chapel,
chamberlain2007parallel}.

UPC++ is a C++ library which offers a PGAS distributed memory programming model
\cite{zheng2014upc++,bachan2017upc++}.
UPC++ has a heavy focus on asynchronous programming that is absent from \bcl{},
including futures, promises, and callbacks.
UPC++'s remote procedure calls can
be used to create more expressive atomic operations, since all RPCs are
executed atomically in UPC++.  However, these operations require interrupting
the remote CPU, and thus have slower throughput than true RDMA atomic memory
operations.  The current version of UPC++ also lacks a library of data
structures, and UPC++ is closely tied to the GASNet communication library, instead
of supporting multiple backends.

DASH is another C++ library that offers a PGAS distributed memory programming
model \cite{Fuerlinger:2016:DASH}.  DASH has a large focus on structured grid
computation, with excellent support for distributed arrays and matrices and an
emphasis on providing each process with fast access to its local portion of the
distributed array.  
While DASH's data
structures are generic, they do not support storing objects with complex types.
DASH is tied to the DART communication
library, which could
potentially offer performance portability through multiple backends, but is
currently limited to MPI for distributed memory communication.

HPX is a task-based runtime system for parallel C++ programs~\cite{kaiser2014hpx}.
It aims to offer
a runtime system for executing standard C++ algorithms efficiently on parallel
systems, including clusters of computers.  Unlike BCL, which is designed to
use coordination-free RDMA communication, HPX's fundamental primitives are remote
procedure calls used to distribute tasks.

STAPL, or the standard adaptive template library, is an STL-like library of
parallel algorithms and data structures for C++
\cite{Tanase:2011:SPC:1941553.1941586}.  STAPL programs are written at a much
higher level of abstraction than \bcl{}, in a functional
style using special higher-order functions such as map, reduce, and for-each which
take lambda functions as arguments.  From this program description, STAPL
generates a hybrid OpenMP and MPI program at compile time.  Some versions of
STAPL also include a runtime which provides load balancing.  The current version
of STAPL is only available in a closed beta and only includes array and vector
data structures \cite{staplguide}.

The Multipol library provided a set of concurrent data structures on top of active
messages, including dynamic load balancing and optimistic task schedulers
\cite{chakrabarti1995multipol}.  However, it was non-portable and did not have
the rich set of hash table data structures discussed here nor the notion of
concurrency promises.

Global Arrays provides a portal shared memory interface, exposing globally
visible array objects that can be read from and written to by each process
\cite{nieplocha1996global}.  While many application-specific libraries have been
built on top of Global Arrays, it lacks the kind of high-level generic data
structures that are the focus of this work.

\section{Conclusion}
\bcl{} is a distributed data structures library that offers productivity
through
high-level, flexible interfaces but maintains high performance by
introducing minimal overhead, offering high-level abstractions that can be
directly compiled down to a small number of one-sided remote memory operations.
We have demonstrated that \bcl{} matches or exceeds the performance of both
hand-optimized domain-specific implementations and general libraries on a range
of benchmarks and is portable to multiple HPC~systems.

\bibliographystyle{plain}
\bibliography{paper}

\clearpage

\end{document}